\documentclass[runningheads]{llncs}

\usepackage[T1]{fontenc}
\def\doi#1{\href{https://doi.org/\detokenize{#1}}{\url{https://doi.org/\detokenize{#1}}}}
\usepackage{graphicx}
\usepackage{listings}
\usepackage[hyphens]{url}
\usepackage{booktabs}
\usepackage{subcaption}
\usepackage{xcolor}

\begin{document}
\title{A Real-time System for Detecting Landslide Reports on Social Media using Artificial Intelligence
}
\titlerunning{A Real-time System for Detecting Landslide Reports on Social Media}
%
\author{Ferda Ofli\inst{1}\orcidID{0000-0003-3918-3230} \and
Umair Qazi\inst{1}\orcidID{0000-0002-2448-9694} \and
Muhammad Imran\inst{1}\orcidID{0000-0001-7882-5502} \and
Julien Roch\inst{2} \and
Catherine Pennington\inst{3}\orcidID{0000-0002-3560-9030} \and
Vanessa Banks\inst{3}\orcidID{0000-0001-6335-7080} \and
Remy Bossu\inst{2,4}\orcidID{0000-0002-9927-9122}
}
\authorrunning{F. Ofli et al.}
%
\institute{
    Qatar Computing Research Institute, Hamad Bin Khalifa University, Doha, Qatar\\
    \email{\{fofli,uqazi,mimran\}@hbku.edu.qa}
\and
    European-Mediterranean Seismological Centre, Arpajon, France\\
    \email{\{julien.roch,bossu\}@emsc-csem.org}
\and
    British Geological Survey, Keyworth, Nottinghamshire, United Kingdom\\
    \email{\{cpoulton,vbanks\}@bgs.ac.uk}
\and
    CEA, DAM, DIF, F-91297 Arpajon, France\\
}
\maketitle              
\begin{abstract}
This paper presents an online system that leverages social media data in real time to identify landslide-related information automatically using state-of-the-art artificial intelligence techniques. The designed system can (i) reduce the information overload by eliminating duplicate and irrelevant content, (ii) identify landslide images, (iii) infer geolocation of the images, and (iv) categorize the user type (organization or person) of the account sharing the information. The system was deployed in February 2020 online at \url{https://landslide-aidr.qcri.org/landslide_system.php} to monitor live Twitter data stream and has been running continuously since then to provide time-critical information to partners such as British Geological Survey and European Mediterranean Seismological Centre. We trust this system can both contribute to harvesting of global landslide data for further research and support global landslide maps to facilitate emergency response and decision making.

\end{abstract}
\section{Introduction}
\label{sec:introduction}

Landslides cause thousands of deaths and billions of dollars in infrastructural damage worldwide every year~\cite{kjekstad2009economic}. However, landslide events are often under-reported and insufficiently documented due to their complex natural phenomena oftentimes triggered by earthquakes and tropical storms, which are more conspicuous, and hence, more widely reported~\cite{lee2004landslide}. Therefore, any attempt to quantify global landslide hazards and the associated impacts remains an underestimation due to this oversight and lack of global data inventories~\cite{froude2018global}.

Undertaking the challenge of building a global landslide inventory, NASA launched a website\footnote{\url{https://gpm.nasa.gov/landslides/index.html}} in 2018 to allow citizens to report about the regional landslides they see in-person or online~\cite{juang2019using}. Following a similar Volunteered Geographical Information (VGI) approach, researchers further developed other means such as mobile or web applications to collect citizen-provided data~\cite{choi2018utilizing,kocaman2019citsci}. 
While VGI-based solutions prove helpful, they are not easily scalable as they require active participation of volunteers that opt-in to use a particular application to collect and share landslide-related data. Furthermore, this means the bulk of data collection and interpretation still involves time consuming work by specialists searching the Internet for news and reports, or directly engaging in communications with those submitting information~\cite{kocaman2019citsci,juang2019using,pennington2015national,taylor2015enriching}.

To alleviate the need for opt-in participation and manual processing, we developed an online system equipped with state-of-the-art AI models to automatically detect landslide reports
posted on social media image streams in real time. The system was developed through an interdisciplinary collaboration between the computer scientists at the Qatar Computing Research Institute (QCRI) and the earthquake and landslide specialists from the European-Mediterranean Seismological Centre (EMSC) and the British Geological Survey (BGS), respectively. 
The developed system employs several supervised machine learning models to (i) deal with the noisy nature of the social media data by filtering out duplicate and irrelevant images, (ii) detect landslide reports by interpreting the retained images, (iii) infer the location information of the detected landslide reports from the available metadata, and (iv) identify the type of users that have shared the landslide reports. We deployed the system online in February 2020 to monitor live Twitter data stream and it has collected more than 54 million tweets and 15 million image URLs. Only about 2.5 million of these image URLs were deemed unique and downloaded for further analysis. Eventually, the system identified about 38,000 landslide reports worldwide, which corresponds to less than 1\% of the collected image URLs and highlights the challenging nature of the problem. 
Despite this, quantitative verification of the system's performance during a real-world deployment shows that our system can detect landslide reports with Precision=76\% and Recall=74\% (i.e., F1=75\%).  


\section{Related Work}
\label{sec:related_work}

The literature on landslide detection and mapping approaches mainly uses four types of data sources: (i) physical sensors, (ii) remote sensing, (iii) volunteers, and (iv) social networks. Sensor-based approaches rely on land characteristics such as rainfall, altitude, soil type, and slope to detect landslides and develop models to predict future events~\cite{merghadi2020machine,ramesh2009wireless}. 
While these approaches can be highly accurate at sub-catchment levels, their large-scale deployment is extremely costly.

Earth observation data from high-resolution satellite imagery has been widely used for landslide detection, mapping, and monitoring~\cite{tofani2013use}. Remote sensing techniques either use Synthetic Aperture Radar (SAR) or optical imagery to perform landslide detection in various formulations including classification, segmentation, object detection, and change detection~\cite{mohan2021review,cheng2013automatic,tavakkoli2019landslide,ji2020landslide,prakash2020mapping,prakash2021new}. While remote sensing through satellites can be useful to monitor landslides globally, their deployment can prove costly and time-consuming. 

A few studies demonstrated the use of Volunteered Geographical Information (VGI) as an alternative method to detect landslides~\cite{choi2018utilizing,kocaman2019citsci,can2019convolutional,can2020development}. These studies assume active participation of volunteers to collect landslide data where the volunteers opt-in to use a mobile or web application to provide information such as photos, time of occurrence, damage description, and other observations about a landslide event. On the contrary, our work capitalizes on massive social media data without any active participation requirement and with better scalability. 

Use of social media data for landslide detection has not been explored extensively. 
The most relevant work by Musaev et al.~\cite{musaev2014litmus,musaev2017rex} combines social media text data and physical sensors to detect landslides. 
In contrast, we focus on analyzing social media images which can provide more detailed information about the impact of the landslide event. To this end, our work complements prior art. 




\section{System Design}
\label{sec:system}

\begin{figure}[t]
    \centering
    \includegraphics[width=0.8\columnwidth]{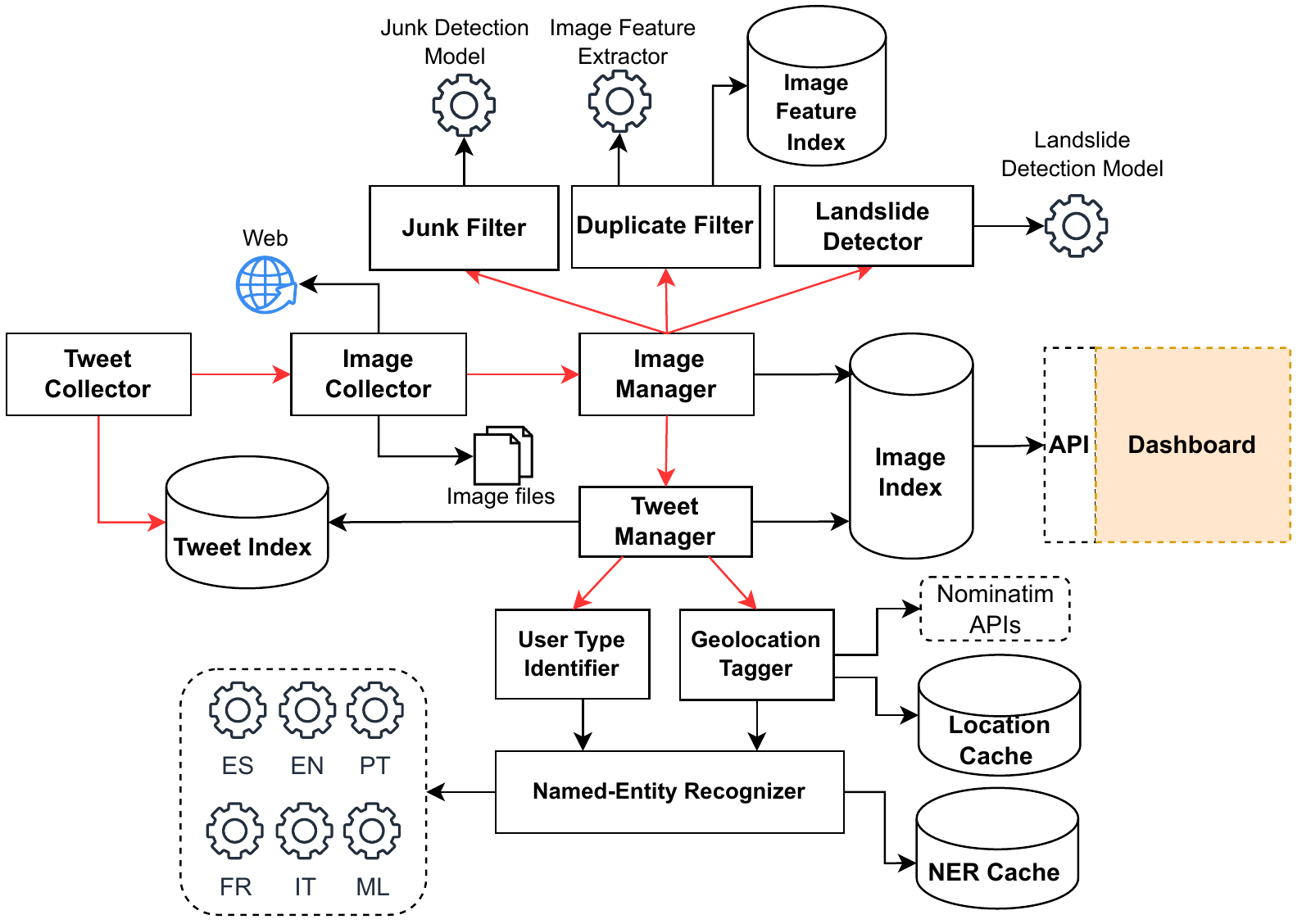}
    \caption{System architecture with important components and communication flows}
    \label{fig:system_diagram}
\end{figure}

The system is designed to ingest data from an online social media platform (i.e., Twitter), analyze the incoming data, and process relevant information under the condition that all tasks must be performed in a time-sensitive manner. Fig.~\ref{fig:system_diagram} shows a high-level architecture of the system and its various critical components. Data flows from left to right through two types of connections between components. The red lines indicate streaming connections whereas the black lines represent on-demand connections. A streaming connection can be of two types (i) a publisher-subscriber channel, and (ii) a push-pop queue. 

\subsection{Data Collectors}\label{ssec:sys_data_collectors}

We have two types of collectors. One collects data (i.e., tweets) directly from Twitter. The other one then downloads images corresponding to collected tweets.

\subsubsection{Tweet Collector}\label{sssec:sys_tweet_collector}
This module uses the Twitter Streaming API\footnote{\url{https://developer.twitter.com/en/docs/twitter-api/v1/tweets/filter-realtime/guides/connecting}} to collect live tweets. The Streaming API can provide data in various ways based on (i) a list of keywords, (ii) geographical bounding boxes, or (iii) both. Our system employs only the keyword-based data collection approach since the bounding box approach provides only geo-tagged tweets which can be about any topic. Tweets matching with at least one of the pre-specified keywords are acquired from Twitter in the JSON format and persisted into the Tweet Index, which is an Elasticsearch database. If a tweet contains one or more images, its \textit{id} and \textit{URLs} of all images are pushed to the Image Collector through a Redis\footnote{\url{https://redis.io/}} queue.

\subsubsection{Image Collector}\label{sssec:sys_image_collector}
This module parses image-related attributes dispatched by the Tweet Collector module and extracts image URLs and downloads corresponding images. Due to re-tweets, same image URLs may appear multiple times during the data collection. To avoid redundant downloads, the system keeps track of previously seen image URLs in an in-memory \emph{linked hash map} which has O(1) time complexity for adding and searching an element and O(n) space complexity. The downloaded images are saved on the file system and their paths and tweet ids are pushed to the Image Manager queue for further processing. 

\subsection{Image Manager}\label{ssec:sys_image_processors}
The system has multiple modules that analyze images for different purposes. Two of these modules, namely Duplicate Filter and Junk Filter, are tasked to reduce the data noise by eliminating images that are (i) near-or-exact duplicate and (ii) irrelevant for general disaster response, respectively. The third module, Landslide Detector, is the core module that interprets each image as landslide or not-landslide. All image processor modules are managed by the Image Manager, which pops items from its queue and immediately dispatches to the three image processors (i.e., Junk Filter, Duplicate Filter, and Landslide Detector) through their respective queues. The Image Manager also monitors the output of all image processors to 
persist them into the main Image Index.

\subsubsection{Duplicate Filter}\label{sssec:sys_duplicate_filter}
Image-level deduplication is important to discard near-or-exact duplicate images that are often due to high retweeting activity. This module identifies duplicate images to prevent further processing as well as information overload on end users. The module acquires images from its input queue and checks whether a given image is near-or-exact duplicate of previously seen images. To this end, it first extracts features from each image using a deep learning model and then compares these features against an Image Feature Index to detect near-or-exact duplicate cases based on a distance threshold. The image feature index keeps a record of all unique image features. If the module identifies a near-or-exact duplicate, it returns the reference image's id and the computed distance. Otherwise, it tags the image as ``not-duplicate''. 
If the image is ``not-duplicate'', then it is also inserted into the Image Feature Index. Section~\ref{ssec:exp_model_duplicate} presents details of the feature extracting model.

\subsubsection{Junk Filter}\label{sssec:sys_junk_filter}
Even though filtered through landslide-related keywords, the Twitter image stream carries images not pertaining to landslide incidents. Identifying these junk content is important to reduce information overload on end users. To this end, the Junk Filter module pops images from the input queue and processes them through the junk detection model, which outputs a class label (``relevant'' or ``not-relevant'') and a confidence score. More detailed information about the junk detection model is presented in Section~\ref{ssec:exp_model_junk}. The processed images are pushed into the output queue of the module.

\subsubsection{Landslide Detector}\label{sssec:sys_landslide_detector}
As the main objective of the system is to identify images showing landslide incidents, in this module we perform this task using a deep learning computer vision model. The module first acquires images from its input queue and passes them through the landslide classifier, which outputs a class label (``landslide'' or ``not-landslide'') and a confidence score. The landslide classifier is a deep learning image classification model that is presented in detail in Section~\ref{ssec:exp_model_landslide}. The classified images are pushed into the module's output queue.


\subsection{Tweet Manager}\label{ssec:sys_text_processors}
The system contains three modules, namely Geolocation Tagger, User Type Identifier, and Named-Entity Recognizer, that process textual content for different purposes. Specifically, Geolocation Tagger analyzes various tweet metadata fields to infer geolocation information while User Type Identifier focuses on identifying the type of Twitter account. Both modules use Named-Entity Recognizer to tag text tokens with named-entities.

\subsubsection{Geolocation Tagger}\label{sssec:sys_geolocation_tagger}
Identifying the location of landslide incidents reported on Twitter is an important task. A tweet reporting a landslide with some image content may or may not have an explicit mention of the location in the text where the incident took place. In that case, other meta-data fields are examined to find location cues. These fields include, \emph{GPS-coordinates, place, user location}, and \emph{user profile description}. To this end, we use our geolocation tagging approach presented in~\cite{imran2022tbcov} with a different field priority order. We observed that most tweets with landslide reporting images contain location cues in their text content. Therefore, if a tweet does not contain GPS-coordinates, we give high priority to the location names mentioned in the text. Place, user location, and user profile description come later in the order, respectively. The geolocation tagger uses the named-entity recognizer to get named-entities for tweet text and user profile description fields. The geolocation tagger uses Nominatim geocoding and reverse geocoding APIs and tags each tweet with country, state, county, and city information, when possible. More details of the geotagging approach can be found in~\cite{imran2022tbcov}. The module maintains a cache of processed locations to increase its efficiency for recurring requests.

\subsubsection{User Type Identifier}\label{sssec:sys_user_type_classifier}
This module uses the name of the tweet author to determine whether the account is of type person or organization. Landslide incidents reported by personal accounts are more important for our end users than those reported by organizational accounts. For this purpose, we use the English NER model through the Named-Entity Recognizer module, which tags name tokens with one of the several predefined named-entities, including PERSON. 

\subsubsection{Named-Entity Recognizer}\label{sssec:sys_ner_classifier}
As described above, both Geolocation Tagger and User Type Identifier modules use Named-Entity Recognizer to perform their operation. To support these operations for multilingual tweets, we use five NER models representing five international languages, including English, French, Spanish, Portuguese, and Italian. Additionally, we use a multilingual NER model (denoted as ML) for all other languages. All of these multilingual models are publicly available at spaCy\footnote{\url{https://spacy.io/usage/models}}. This module also maintains a cache of processed NER requests to increase its efficiency for recurring requests.




\section{Experiments}
\label{sec:experiments}

In this section, we first describe the design and development of our image models and present experimental results. Then, we present performance evaluation and benchmarking results for the most critical components of the system. For image models, we follow the popular transfer learning approach based on convolutional neural networks (CNNs) as many studies have shown that features learned by CNNs are effectively transferable between different visual recognition tasks~\cite{JDonahue:ICML14,Sermanet:ICLR14,MOquab:CVPR14}, particularly when training samples are limited.

\begin{figure}[t]
    \centering
    \begin{minipage}{\columnwidth}
        \begin{subfigure}[b]{0.48\columnwidth}
            \centering
            \includegraphics[width=.85\columnwidth]{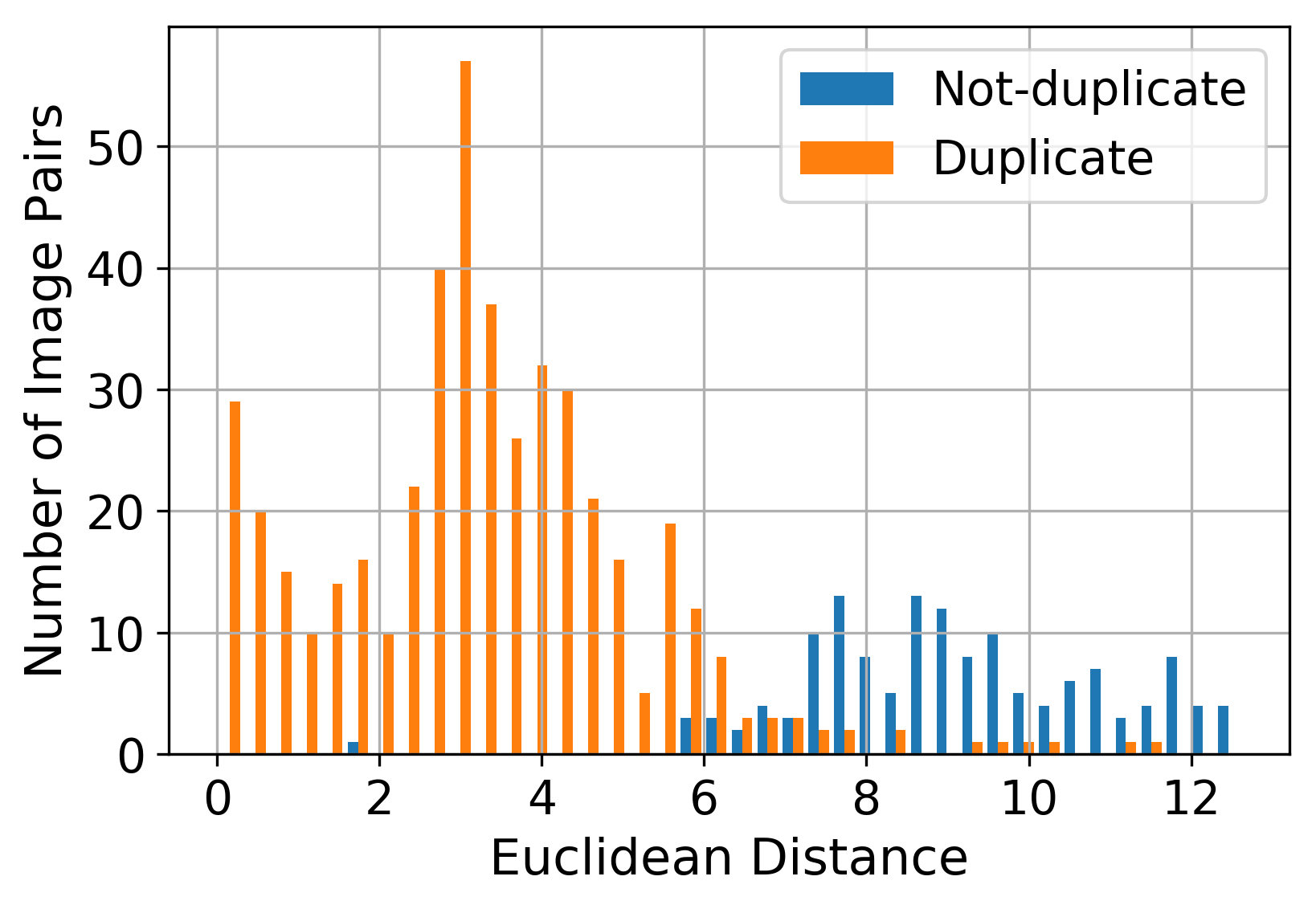}
            \caption[Histogram of pairwise distances]{Histogram of pairwise distances}
            \label{fig:dup_dist_distribution}
        \end{subfigure}
        \hfill
        \begin{subfigure}[b]{0.48\columnwidth}  
            \centering 
            \includegraphics[width=.85\columnwidth]{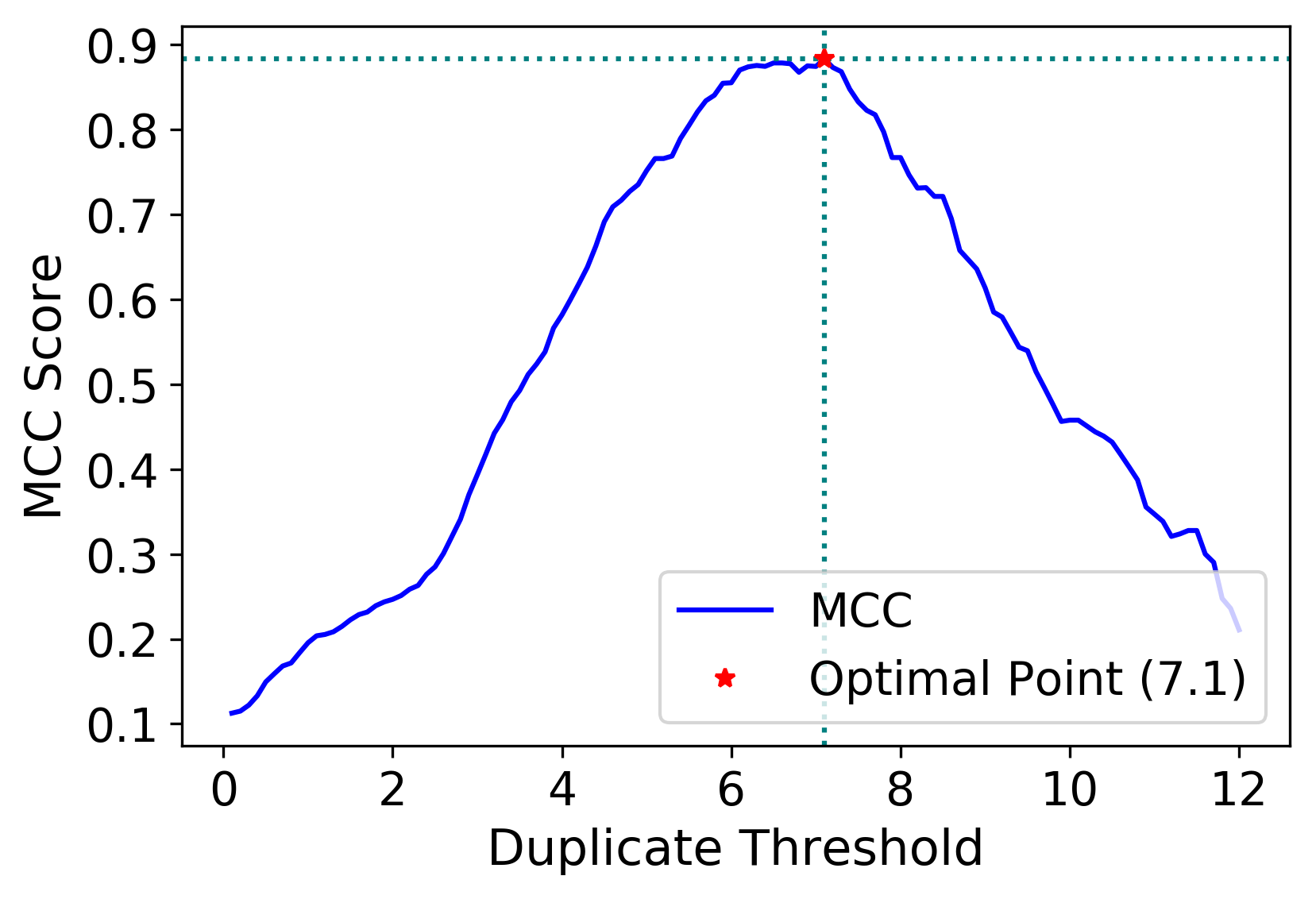}
            \caption[MCC vs.\ threshold]{MCC vs.\ duplicate threshold}
            \label{fig:dup_mcc_curve}
        \end{subfigure}
    \end{minipage}%
    \caption[Optimal duplicate distance threshold determination: (a) Distribution of the Euclidean distances between the image pairs in the duplicate test set. (b) MCC performance as a function of distance threshold.]
    {Optimal duplicate distance threshold determination: (a) Distribution of the Euclidean distances between the image pairs in the duplicate test set. (b) MCC performance as a function of distance threshold.} 
    \label{fig:duplicate_threshold}
    \vspace{-2mm}
\end{figure}

\subsection{Duplicate Filtering}\label{ssec:exp_model_duplicate}
The Duplicate Filter is responsible for extracting a feature vector from a given image using a state-of-the-art deep learning model and comparing this feature vector with the feature vectors of previously seen images based on a pre-defined distance threshold $d$. For this purpose, we extract deep features from the penultimate layer of a ResNet-50 model~\cite{he2016deep} pre-trained on the Places data set~\cite{zhou2017places}, which comprises 10 million images collected for scene recognition.\footnote{The pre-trained model is available at \url{http://places2.csail.mit.edu/models_places365/resnet50_places365.pth.tar} (accessed on Jan 23, 2022).} Each feature vector has a size of 2,048. To determine the optimal distance threshold $d$, we performed experiments on a manually annotated set of 600 image pairs including 460 duplicate and 140 non-duplicate cases with varying pairwise distances (Fig.~\ref{fig:dup_dist_distribution}). We used Euclidean distance metric (i.e., L2 norm) to measure the distance between two image feature vectors. 
Note that image pairs with a distance greater than 12.5 looked trivially distinct, and hence, we did not include them in our experiments. We then performed a grid search over a range of threshold values from 0 to 12 with a step size of 0.1 and measured the performance of each threshold value by computing the Matthew's Correlation Coefficient (MCC), which is regarded as a balanced measure for imbalanced classification problems~\cite{chicco2020advantages}. As depicted in Fig.~\ref{fig:dup_mcc_curve}, the optimal performance is achieved when the duplicate distance threshold is $d=7.1$.

\subsection{Junk Classification}\label{ssec:exp_model_junk}
\begin{figure}[t]
    \centering
    \begin{minipage}{\columnwidth}
        \begin{subfigure}[b]{0.48\columnwidth}
            \centering
            \includegraphics[width=.9\columnwidth]{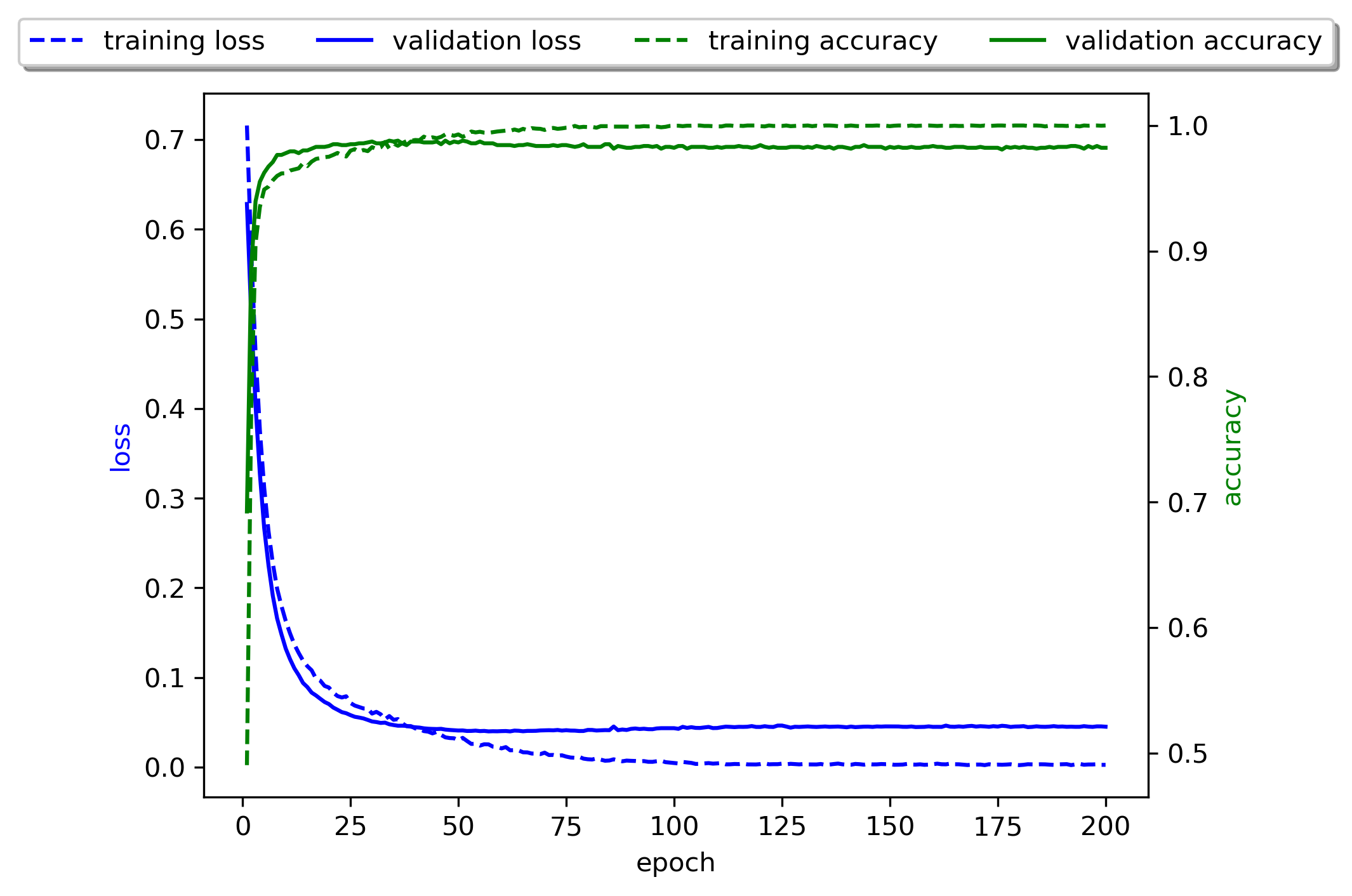}
            \caption[Junk classification]{Junk classification}
            \label{fig:train_prog_relevancy}
        \end{subfigure}
        \hfill
        \begin{subfigure}[b]{0.48\columnwidth}  
            \centering 
            \includegraphics[width=.9\columnwidth]{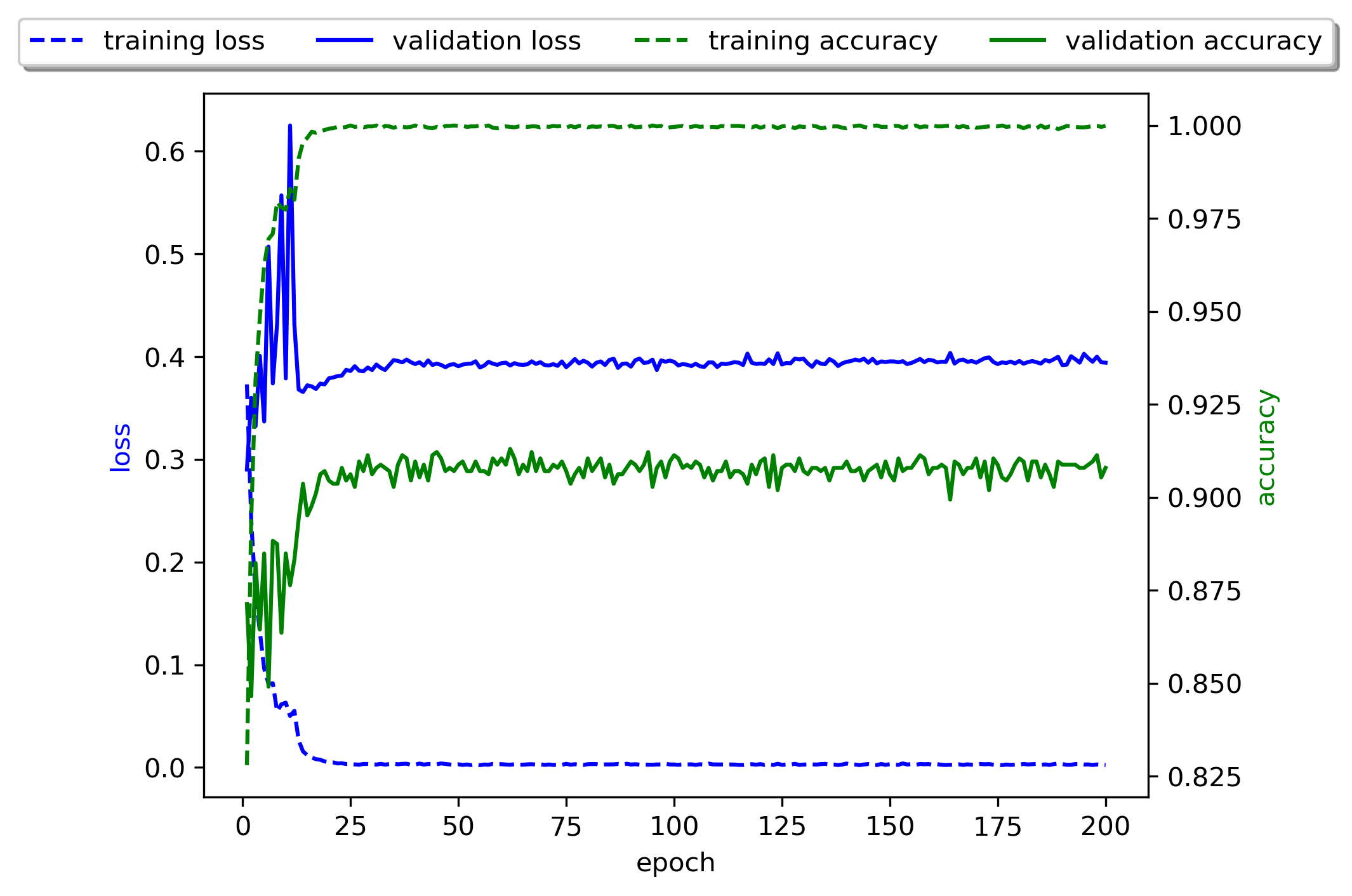}
            \caption[Landslide classification]{Landslide classification}
            \label{fig:train_prog_landslide}
        \end{subfigure}
    \end{minipage}%
    \caption[Model training progress in terms of accuracy and loss achieved on the training and validation sets]
    {Model training progress in terms of accuracy and loss achieved on the training and validation sets} 
    \label{fig:training_progress}
\end{figure}

The Junk Filter employs a CNN model to determine whether an image is relevant or not for general emergency management and response. To this end, we took a ResNet-50 model~\cite{he2016deep} pre-trained on ImageNet~\cite{ILSVRC15}, adopted its final layer to binary classification task, and fine-tuned it on a custom data set introduced by Nguyen et al.~\cite{DTNguyen:ISCRAM17}. We merged the validation set with the training set, and used the test set to evaluate the performance of the model as summarized in Table~\ref{tab:relevancy_model_data}. We used Adam optimizer~\cite{kingma2014adam} with an initial learning rate of $10^{-6}$ and configured the \texttt{ReduceLROnPlateau} scheduler to decay the learning rate by 0.1 with a patience of 50 epochs. We trained the model for a total of 200 epochs. The training process of the junk classification model is plotted in Fig.~\ref{fig:train_prog_relevancy} and its performance evaluation is presented in Table~\ref{tab:relevancy_model_perf}. The model achieves almost perfect performance in all measures due to the distinct features between relevant and not-relevant images in the training data set.

\begin{table}[t]
    \caption[Details of the data set used for training the junk classification model and the performance of the trained model on the test set]{Details of the data set used for training the junk classification model and the performance of the trained model on the test set.}
    \label{tab:relevancy_model}
    \begin{subtable}{0.5\textwidth}
        \centering
        \caption{Training data set}
        \label{tab:relevancy_model_data}
        \begin{tabular}{@{}lrrr@{}}
        \toprule
        \textbf{Class} & \textbf{Train} & \textbf{Test} & \textbf{Total} \\ \midrule
        Relevant      & 2,814   & 704   & 3,518 \\
        Not-relevant  & 2,814   & 704   & 3,518 \\ \midrule
        Total         & 5,628   & 1,408 & 7,036 \\
        \bottomrule
        \end{tabular}
    \end{subtable}
    \hfill
    \begin{subtable}{0.5\textwidth}
        \centering
        \caption{Model performance (Acc: 98.79)}
        \label{tab:relevancy_model_perf}
        \begin{tabular}{@{}lccc@{}}
        \toprule
        \textbf{Class} & \textbf{Precision} & \textbf{Recall} & \textbf{F1} \\ \midrule
        Relevant       & 98.31 & 99.29 & 98.80 \\
        Not-relevant   & 99.28 & 98.30 & 98.79 \\ \midrule
        Macro avg.     & 98.80 & 98.79 & 98.79 \\
        \bottomrule
        \end{tabular}
    \end{subtable}%
\end{table}

\subsection{Landslide Classification}\label{ssec:exp_model_landslide}
The Landslide Detector is the most important component of the proposed system. Therefore, we performed a separate, comprehensive study to identify the optimal configuration for the landslide classification model~\cite{ofli2021landslide}. To recap, we first created a large landslide image data set labeled by landslide specialists, who are also co-authors of this paper. The data set contains 11,737 images, which are split into training, validation, and test sets as shown in Table~\ref{tab:dataset_landslide}. Then, adopting a transfer learning approach, we conducted an extensive set of experiments using various CNN architectures with different optimizers, learning rates, weight decays, and class balancing strategies. The winning model configuration is a ResNet-50 architecture trained using Adam optimizer with an initial learning rate of $10{-4}$, a weight decay of $10^{-3}$, and without a class balancing strategy. Fig.~\ref{fig:train_prog_landslide} displays the training progress of the best performing landslide classification model, which is also integrated into our system, whereas Table~\ref{tab:perf_model_landslide} summarizes the performance of the model on the test set.

\begin{table}[t]
    \caption[Details of the data set used for training the landslide classification model and the performance of the trained model on the test set]{Details of the data set used for training the landslide classification model and the performance of the trained model on the test set.}
    \label{tab:perf_models}
    \begin{subtable}{0.5\textwidth}
        \centering
        \caption{Training data set}
        \label{tab:dataset_landslide}
        \begin{tabular}{@{}lrrrr@{}}
        \toprule
        \textbf{Class} & \textbf{Train} & \textbf{Val} & \textbf{Test} & \textbf{Total} \\ \midrule
        Landslide & 1,883 & 271 & 536 & 2,690 \\
        Not-landslide & 6,332 & 902 & 1,813 & 9,047 \\ \midrule
        Total & 8,215 & 1,173 & 2,349 & 11,737 \\
        \bottomrule
        \end{tabular}
    \end{subtable}
    \hfill
    \begin{subtable}{0.5\textwidth}
        \centering
        \caption{Model performance (Acc: 86.97)}
        \label{tab:perf_model_landslide}
        \begin{tabular}{@{}lccc@{}}
        \toprule
        \textbf{Class} & \textbf{Precision} & \textbf{Recall} & \textbf{F1} \\ \midrule
        Landslide       & 73.66 & 66.79 & 70.06 \\
        Not-landslide   & 90.45 & 92.94 & 91.68 \\ \midrule
        Macro avg.      & 82.05 & 79.87 & 80.87 \\
        \bottomrule
        \end{tabular}
    \end{subtable}%
\end{table}



\subsection{Performance Evaluation and Benchmarking}
\label{ssec:sys_perf}

\begin{figure}[t]
    \centering
    \begin{minipage}{\columnwidth}
        \includegraphics[width=0.24\columnwidth]{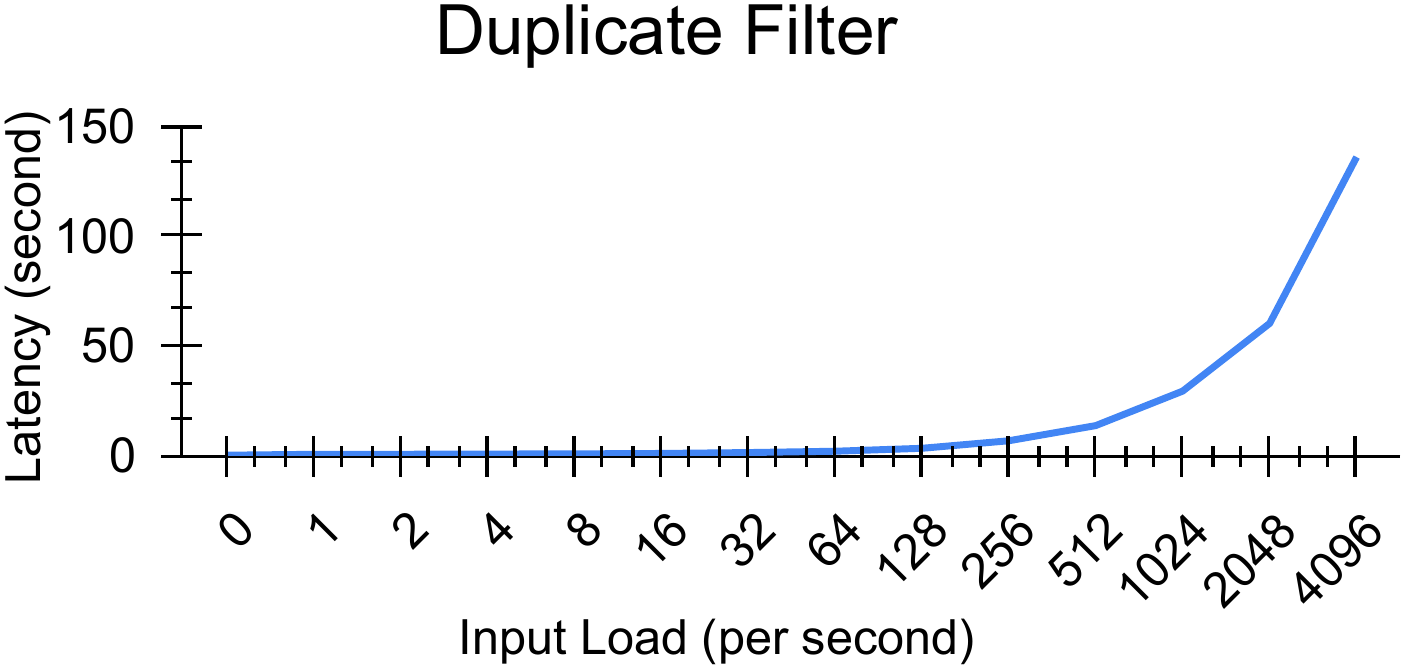}
        \hfill
        \includegraphics[width=0.24\columnwidth]{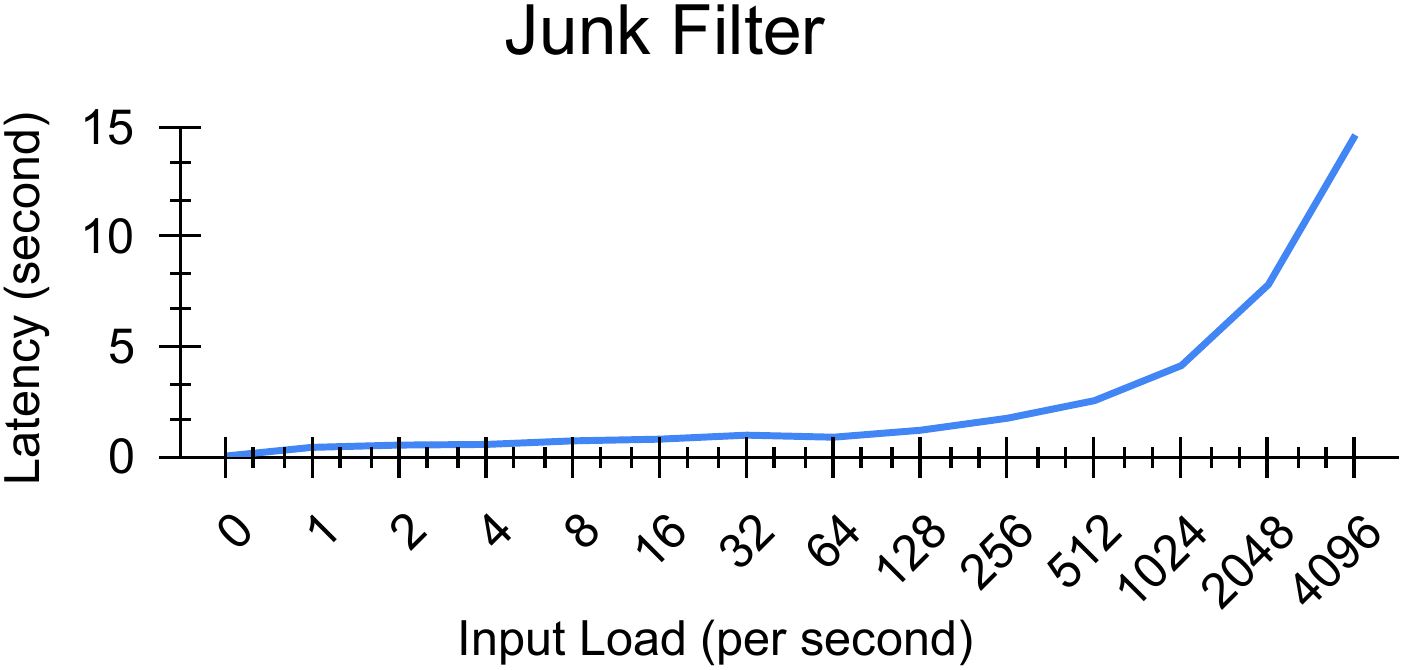}
        \hfill
        \includegraphics[width=0.24\columnwidth]{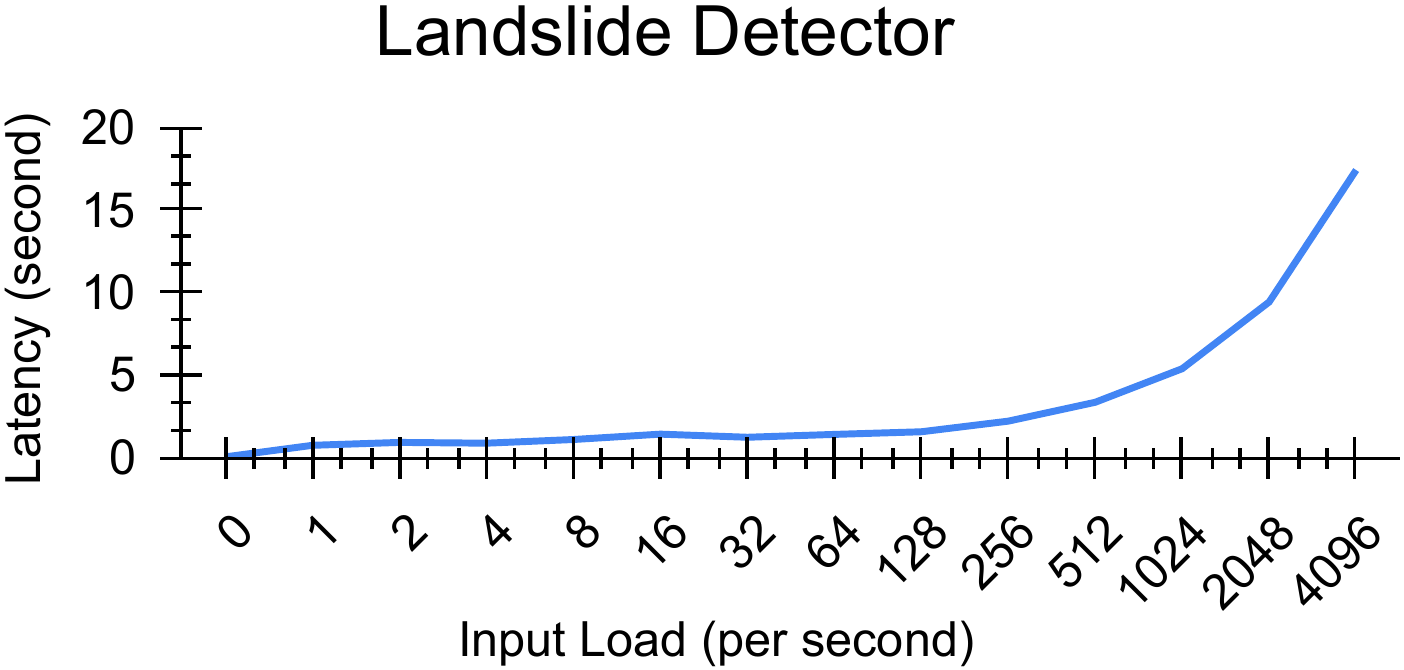}
        \hfill
        \includegraphics[width=0.24\columnwidth]{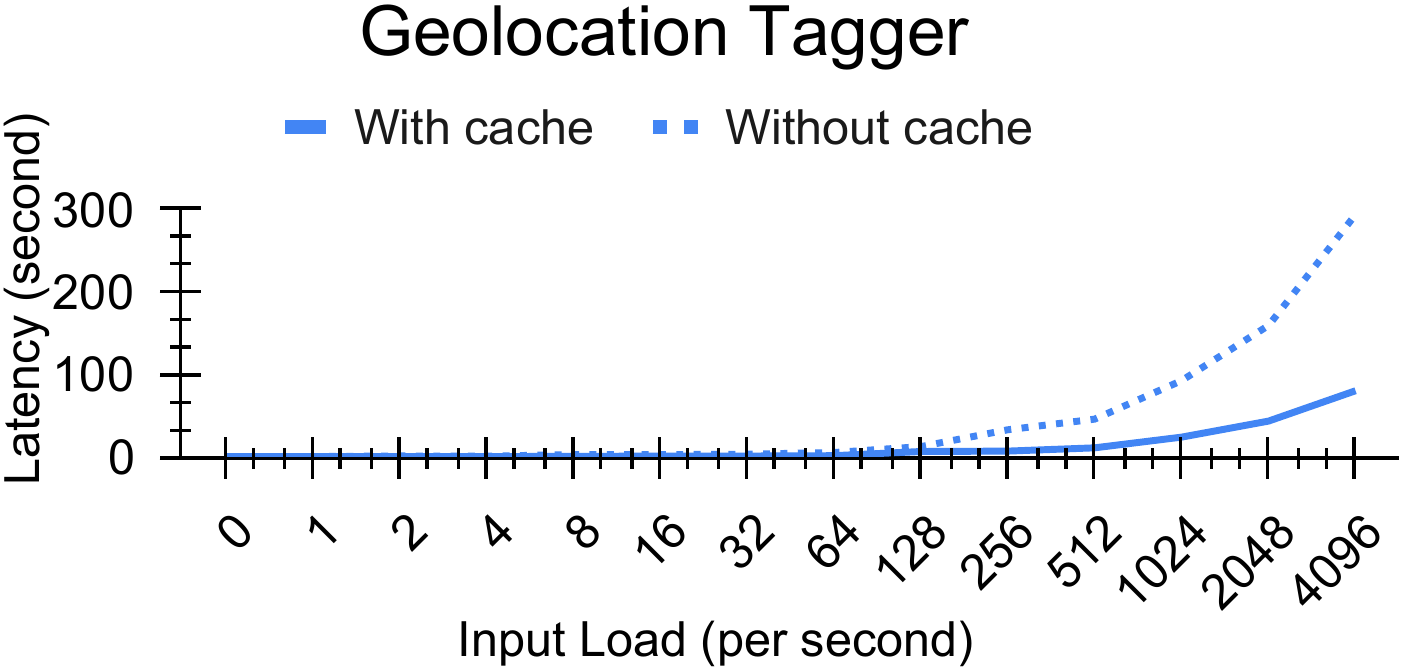}
        \hfill
        \vspace{1mm}
        \includegraphics[width=0.24\columnwidth]{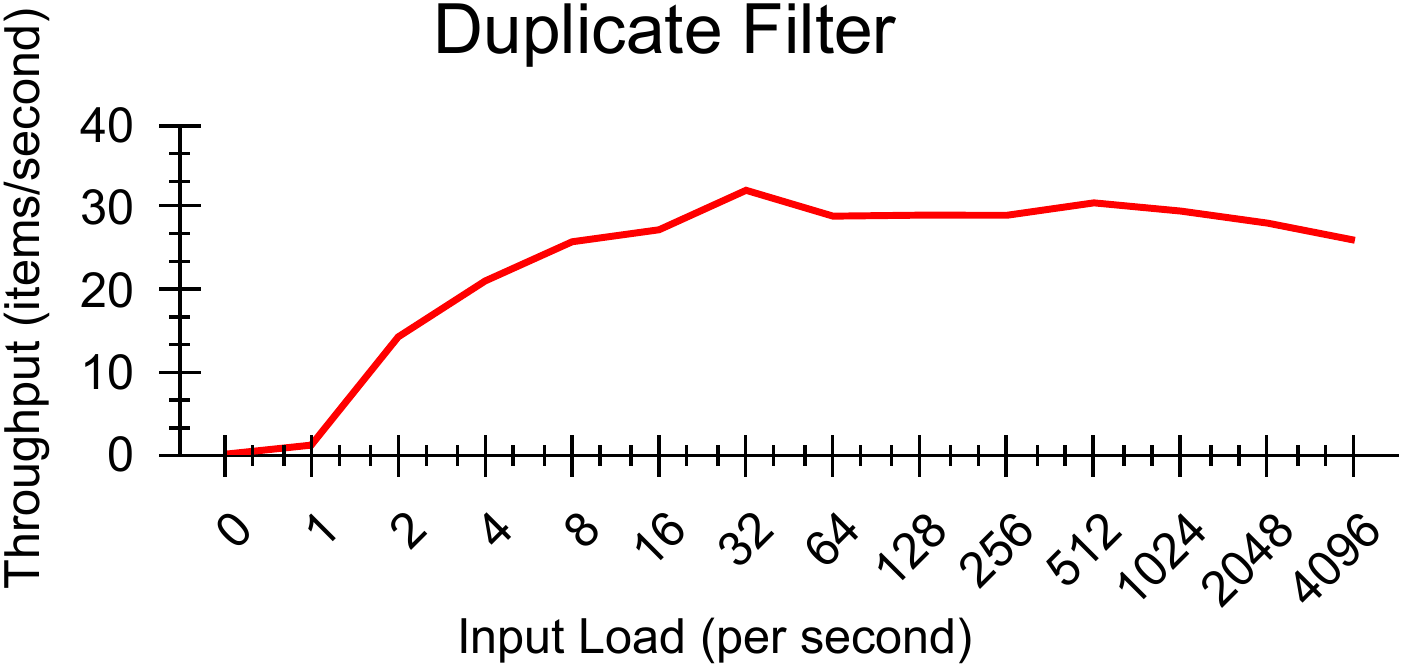}
        \hfill
        \includegraphics[width=0.24\columnwidth]{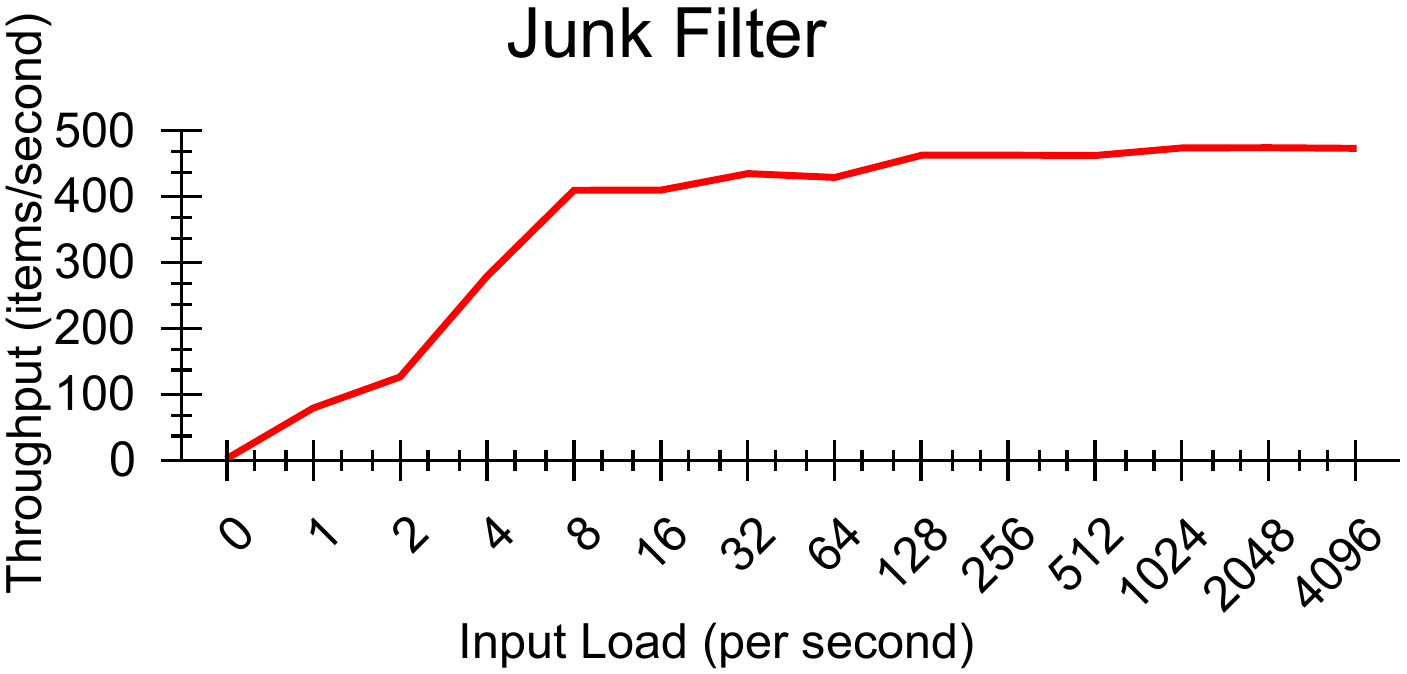}
        \hfill
        \includegraphics[width=0.24\columnwidth]{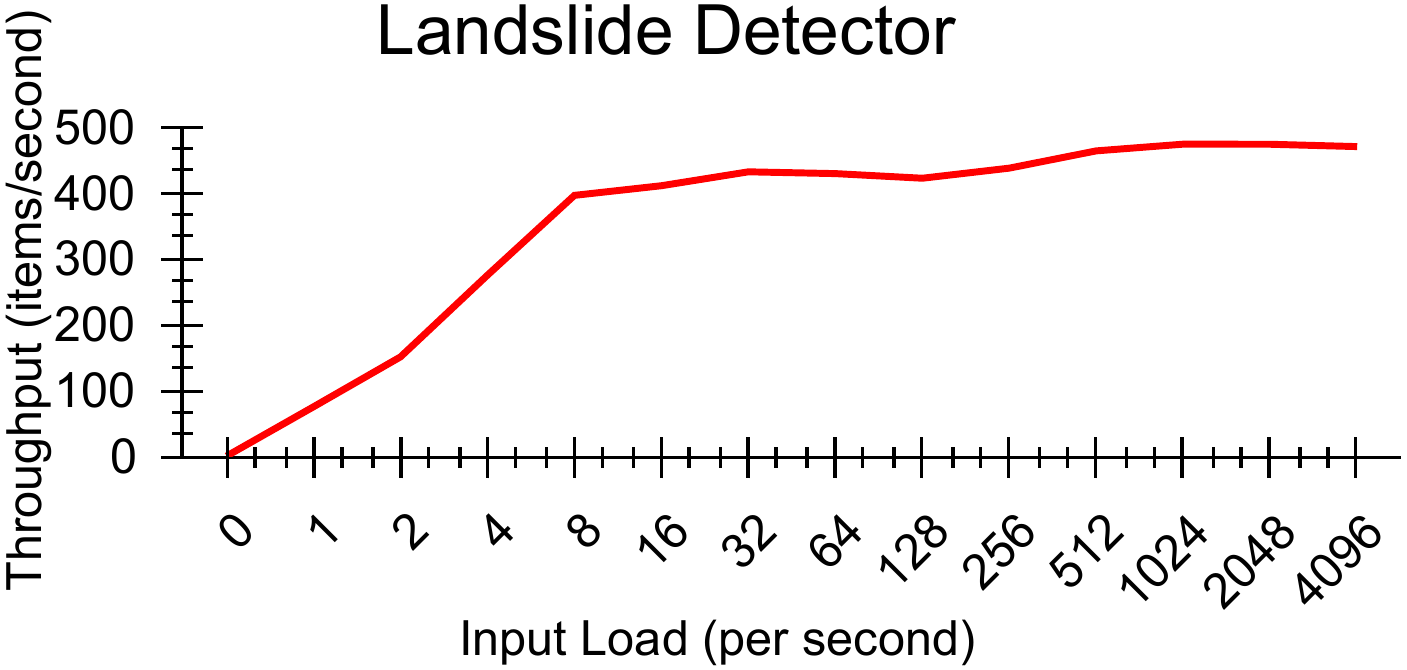}
        \hfill
        \includegraphics[width=0.24\columnwidth]{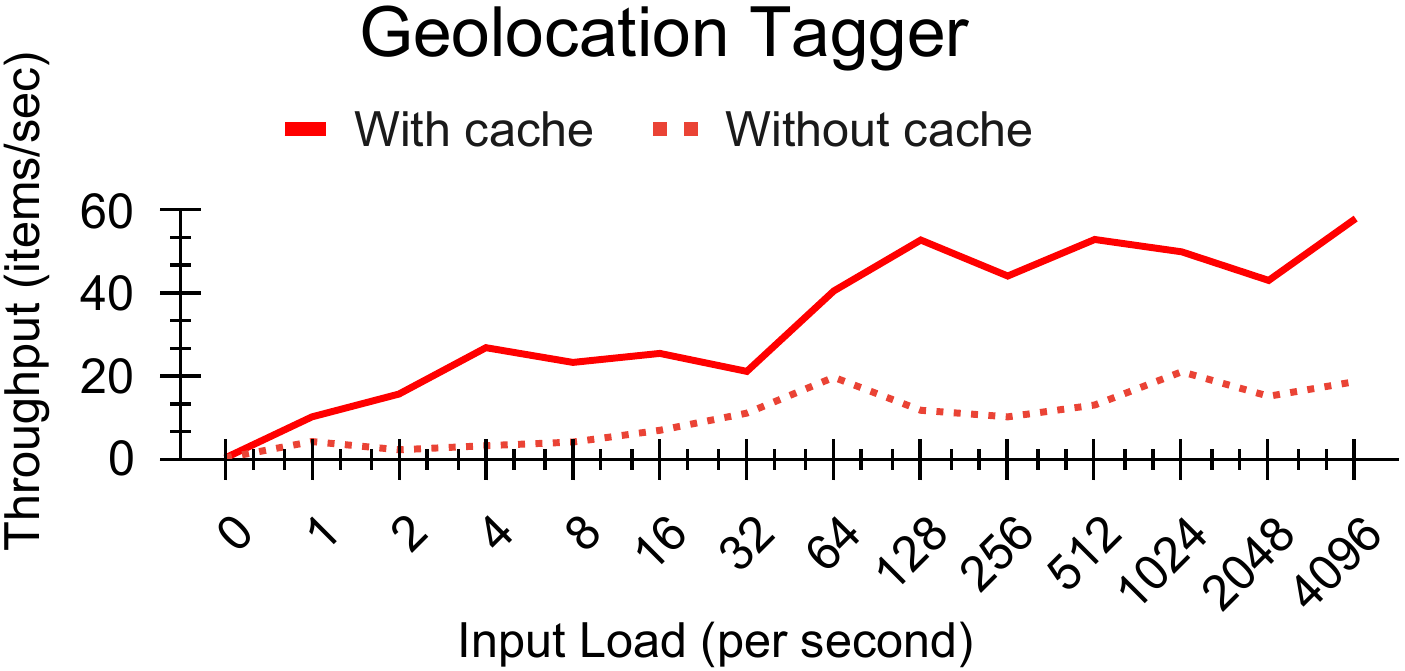}
        \hfill
    \end{minipage}
    \caption[Latency (top) and throughput (bottom) of the Junk Filter, Duplicate Filter, Landslide Detector, and Geolocation Tagger (left to right).]
    {Latency (top) and throughput (bottom) of the Junk Filter, Duplicate Filter, Landslide Detector, and Geolocation Tagger (left to right).} 
    \label{fig:latecy_throughput_alltogether}
\end{figure}


To stress-test the system and understand its scalability, we conducted performance experiments on four critical modules, i.e., Duplicate Filter, Junk Filter, Landslide Detector, and Geolocation Tagger. We use latency and throughput, as they are considered reliable measures to test a system's performance. In our case, the latency is the time taken by a module to process a given input load consisting of images. Whereas, the throughput is the number of images processed in a unit time (one second) given an input load. The experiments were conducted using a pool of ~50,000 images. We developed a simulator to mimic the functionality of the Image Collector. The simulator pushed varying amounts of input loads to Redis channels, which were then consumed by modules. Based on the real-world deployment, we observed that the input load reaches a maximum of 0.08 images per second (on average). Therefore, we tested a range of input loads defined as $2^n, n\in\{0, 1, ..., 12\}$. We performed the tests on a Linux server with 256GB RAM, 2.2 GHz processor with 32 cores and two Tesla V100 GPUs with 16GB. 

Fig.~\ref{fig:latecy_throughput_alltogether} shows the performance results. The latency for all modules follows the same pattern, i.e., as the input load (per second) increases, the latency also increases. However, as the computational responsibilities of each module differ, so do their latencies at different input loads. For instance, both Relevancy Filter and Landslide Detector show a decent latency of around five seconds even at 1024 input load. The Duplicate Filter, however, exhibits high latency (i.e., 29 seconds) at the same load. The latency for Geolocation Tagger is measured with and without cache, which makes a significant difference. The cache keeps a record of all existing unique requests and hence, on average, the latency of the cached version is about four times less.

In terms of throughput, Relevancy Filter and Landslide Detector maintain a high throughput of more than 400 images/second, even at the maximum input load. Throughput for Junk Filter reaches module capacity at 467 images/second on average and for Landslide Detector it goes up to 457 images/second on average. For Duplicate Filter, the throughput initially increases but then starts decreasing as the size of Image Feature Index grows. The throughput is also about 4 times higher on average with cache compared to without cache. Geolocation Tagger reaches its capacity at about 50 images per second with cache. With an empty cache, it goes as high as 21 images per second on average with cache and is expected to increase as the cache grows in size.




\section{Real-world Deployment}
\label{sec:deployment}

Here we present details about our real-world deployment including data collection and statistics, quantitative verification of the detected landslide reports, and a comparison with a text-based approach.

\begin{figure}[t]
\centering
\includegraphics[width=.88\columnwidth]{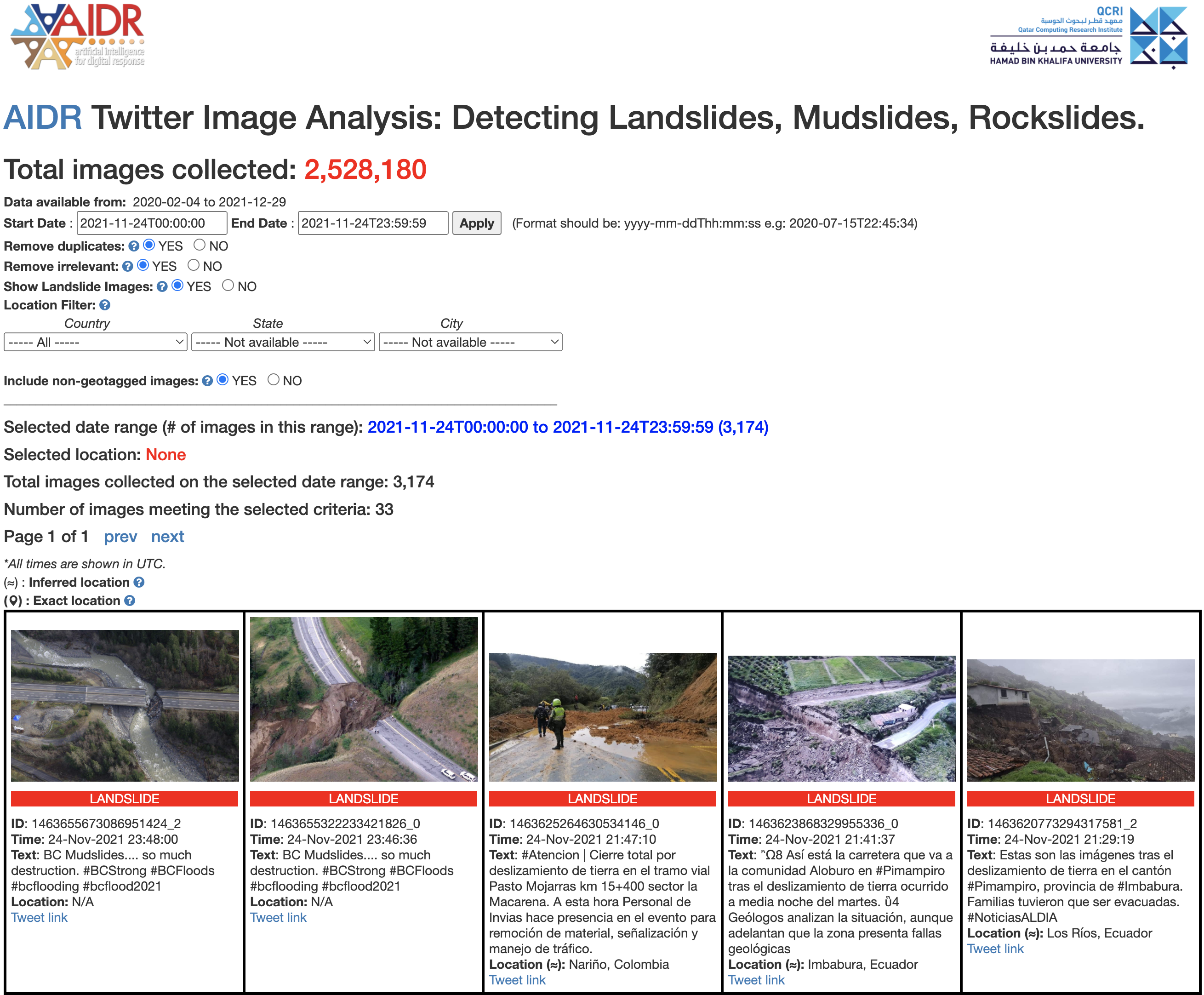}
\caption{Snapshot of the online system}
\label{fig:system_screenshot}
\vspace{-3mm}
\end{figure}

\subsection{Data Collection and Statistics}

In February 2020, we launched the system online at \url{https://landslide-aidr.qcri.org/landslide_system.php} to monitor live Twitter stream for landslide-related reports. Fig.~\ref{fig:system_screenshot} shows a snapshot of the system dashboard. We note that, by landslide, we refer to all downward and outward movement of loosened slope materials such as landslip, debris flows, mudslides, rockfalls, earthflows, and other mass movements. As mentioned in Section~\ref{ssec:sys_data_collectors}, the system follows a keyword-based data collection strategy. Hence, we curated a list of 339 multilingual keywords covering all types of landslides in 32 languages including English, Albanian, Arabic, Basque, Bengali, Bosnian, Catalan, Chinese, Croatian, Dutch, French, Georgian, German, Greek, Hindi, Hungarian, Indonesia, Iranian, Italian, Japanese, Korean, Malaysia, Philippines, Polish, Portuguese, Romanian, Russian, Sesotho, Slovenian, Spanish, Swedish, and Turkish (Table~\ref{tab:collection_terms}).

\begin{table}[t]
\caption{List of all keywords in 32 languages used for data collection.}
\label{tab:collection_terms}
\centering
\begin{tabular}{c}
\toprule
\includegraphics[width=.73\columnwidth]{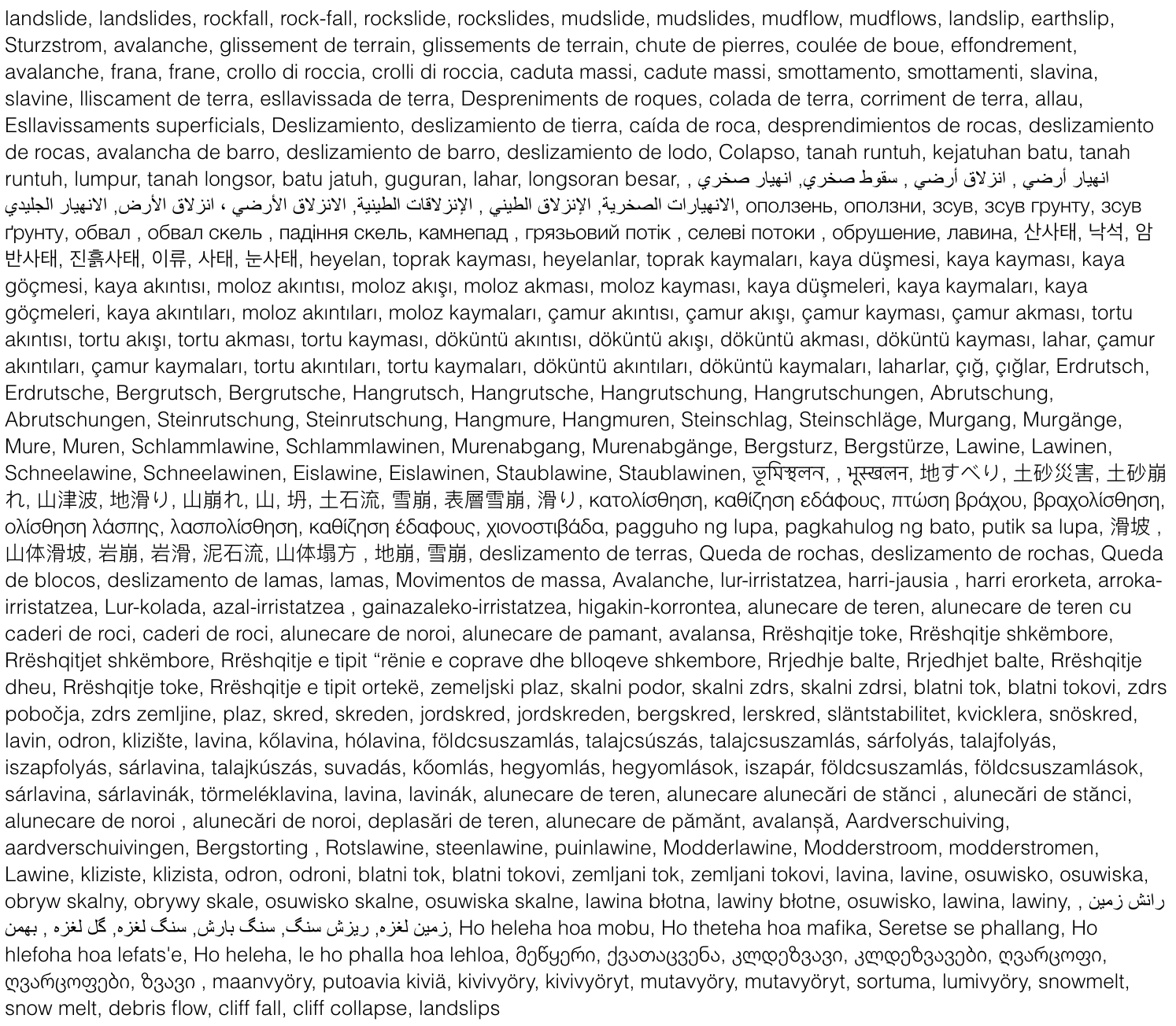}\\
\bottomrule
\end{tabular}
\vspace{-5mm}
\end{table}

\begin{figure}[t]
\centering
\includegraphics[width=\linewidth]{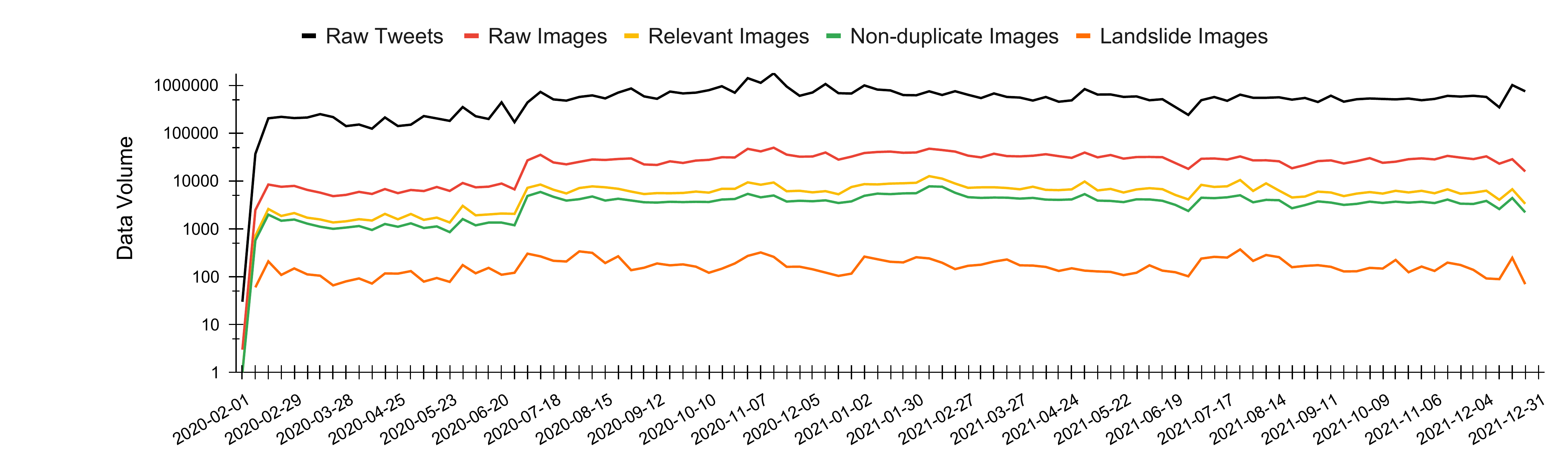}
\caption{Weekly distributions of raw tweets and images as well as the relevant, non-duplicate, and landslide images (y-axis is in log scale).}
\label{fig:weekly_dist_raw_filt}
\end{figure}

\begin{figure}[t]
    \centering
    \begin{minipage}{\columnwidth} 
        \centering
        \includegraphics[width=0.49\columnwidth]{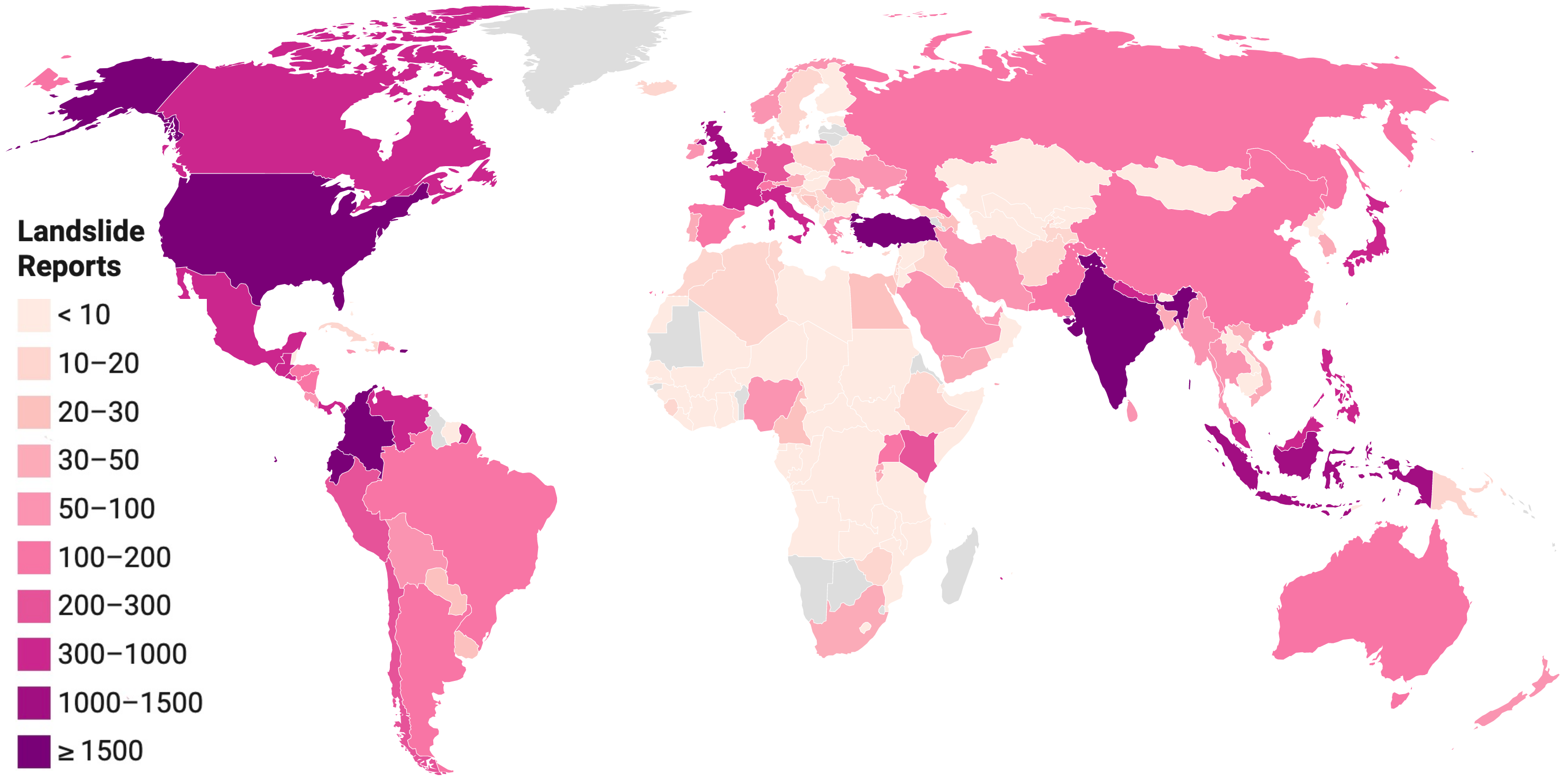}
        \hfill
        \includegraphics[width=0.49\columnwidth]{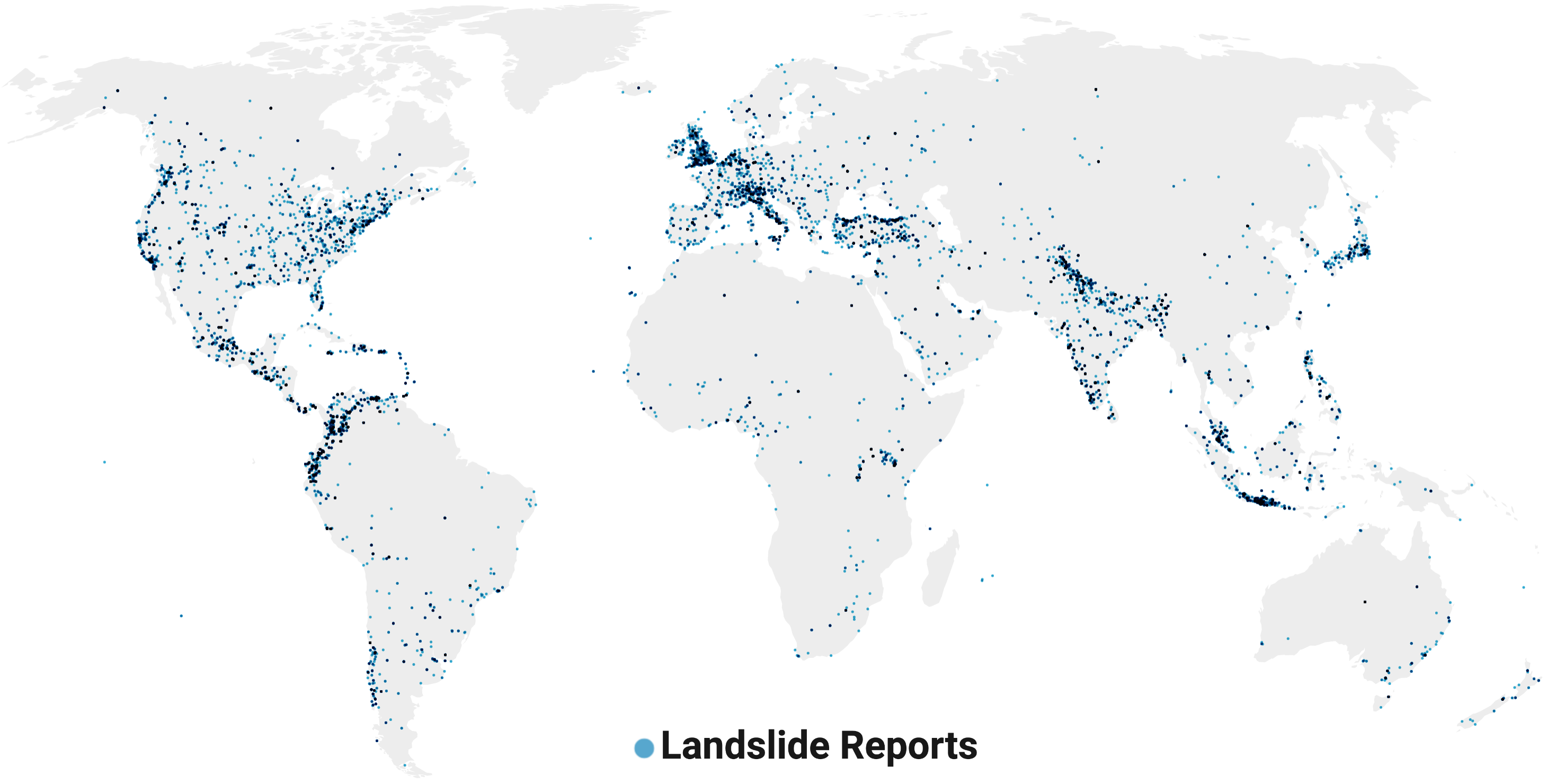}
        \hfill
    \end{minipage}
    \caption[Worldwide distribution of the collected landslide reports]
    {Worldwide distribution of the collected landslide reports}
    \label{fig:global_map}
    \vspace{-3mm}
\end{figure}

\begin{figure}[t]
    \centering
    \begin{minipage}{\columnwidth} 
        \centering
        \includegraphics[width=0.24\columnwidth]{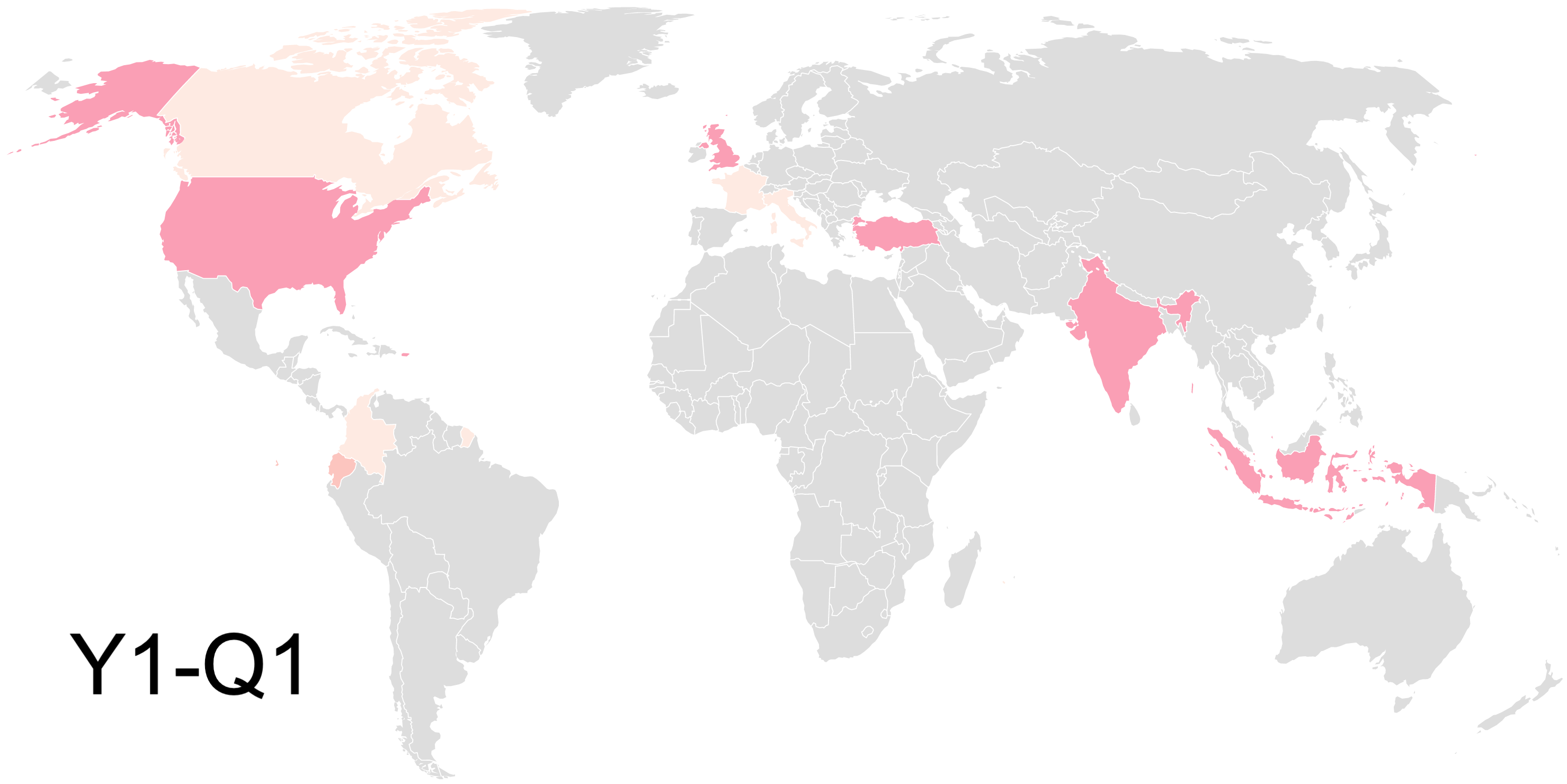}
        \hfill
        \includegraphics[width=0.24\columnwidth]{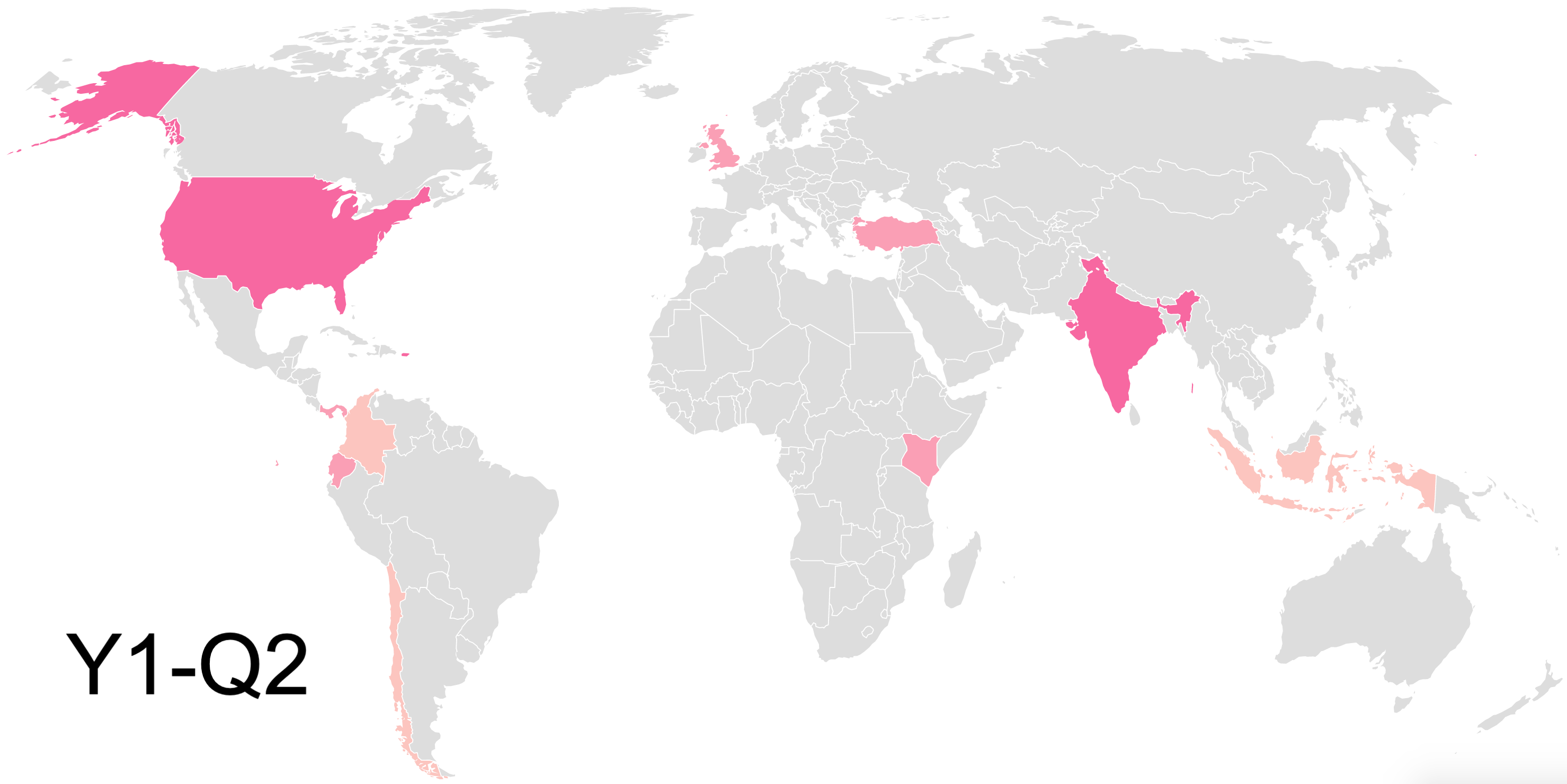}
        \hfill
        \includegraphics[width=0.24\columnwidth]{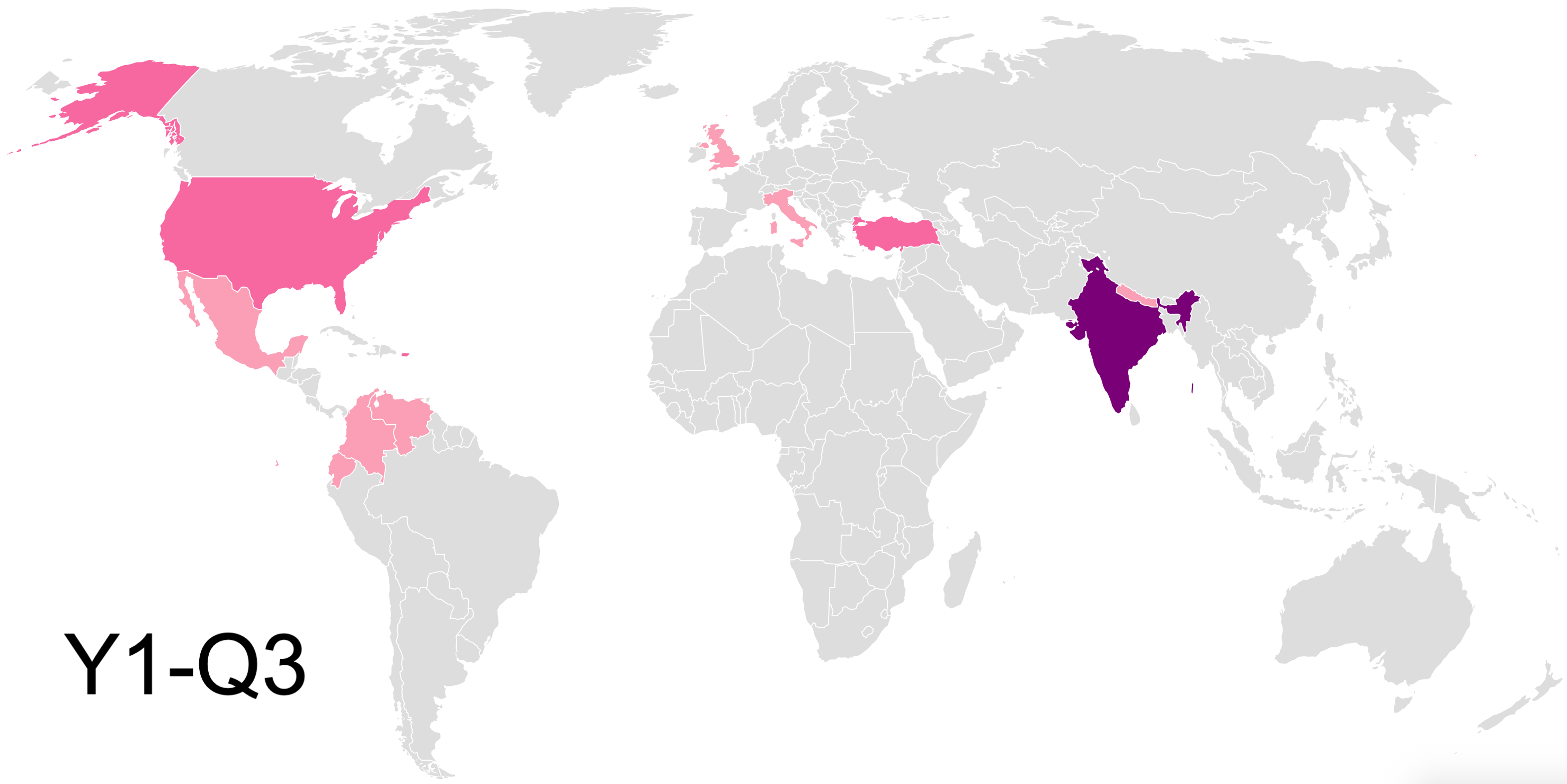}
        \hfill
        \includegraphics[width=0.24\columnwidth]{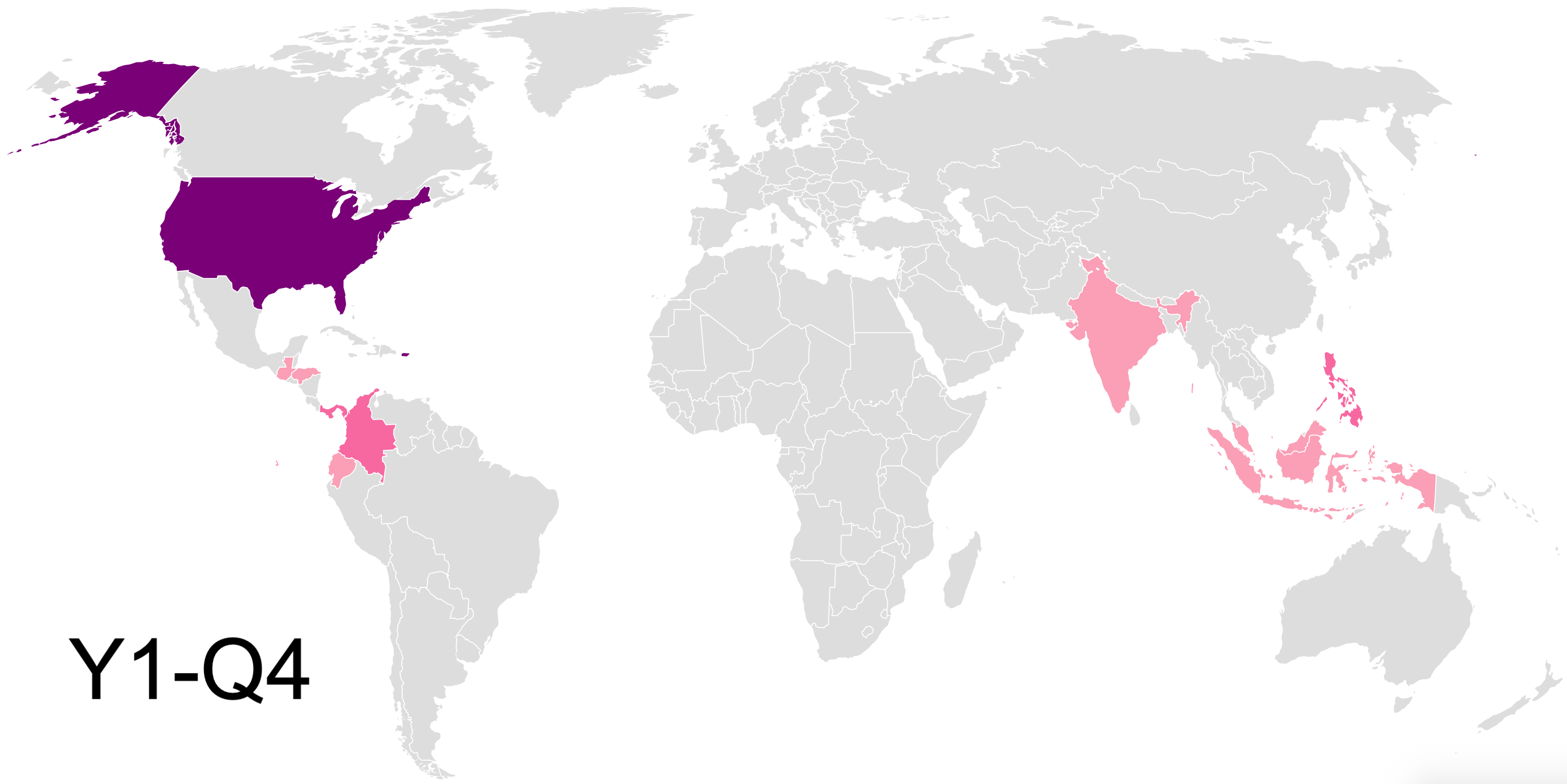}
        \hfill
        \includegraphics[width=0.24\columnwidth]{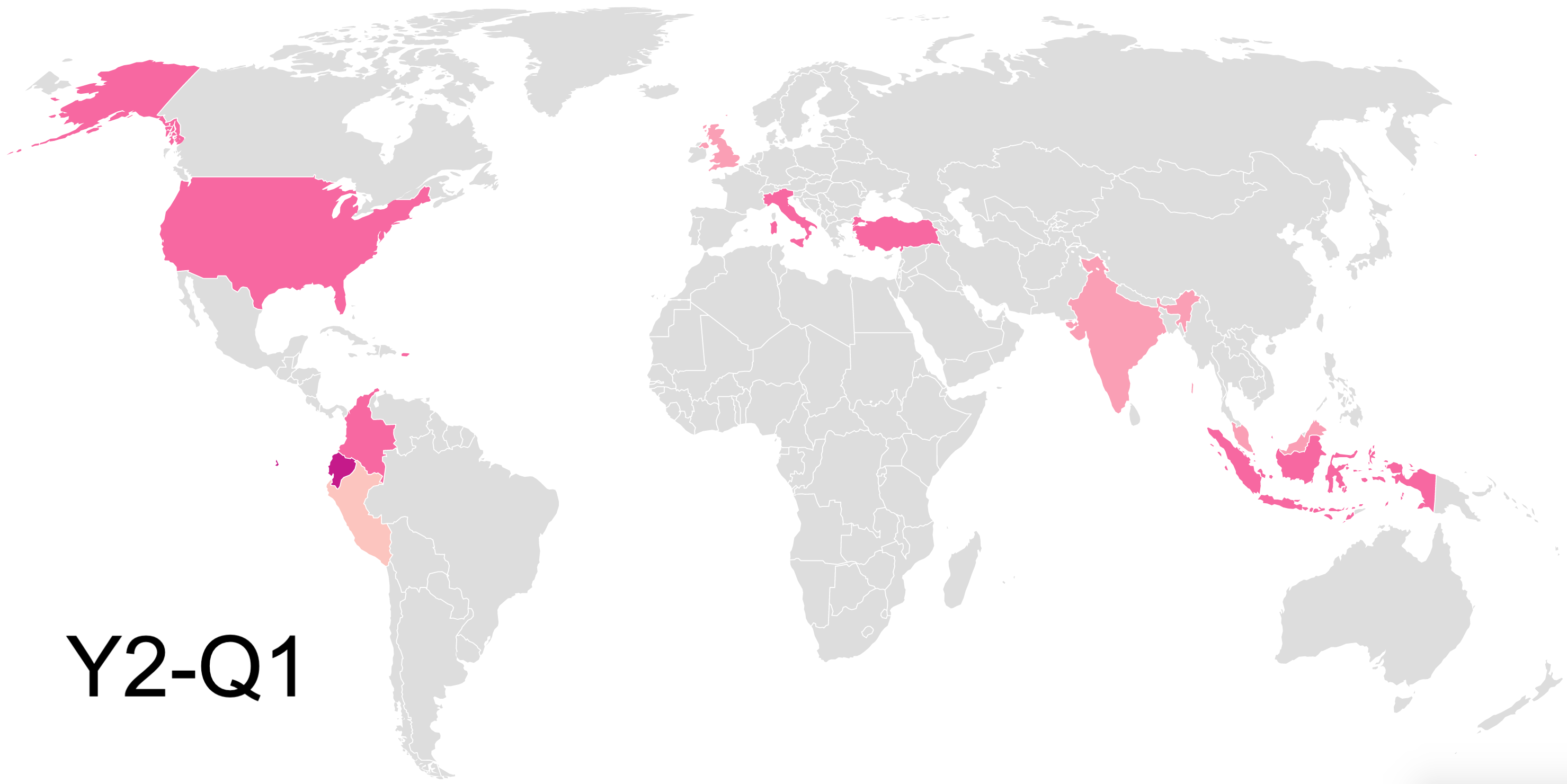}
        \hfill
        \includegraphics[width=0.24\columnwidth]{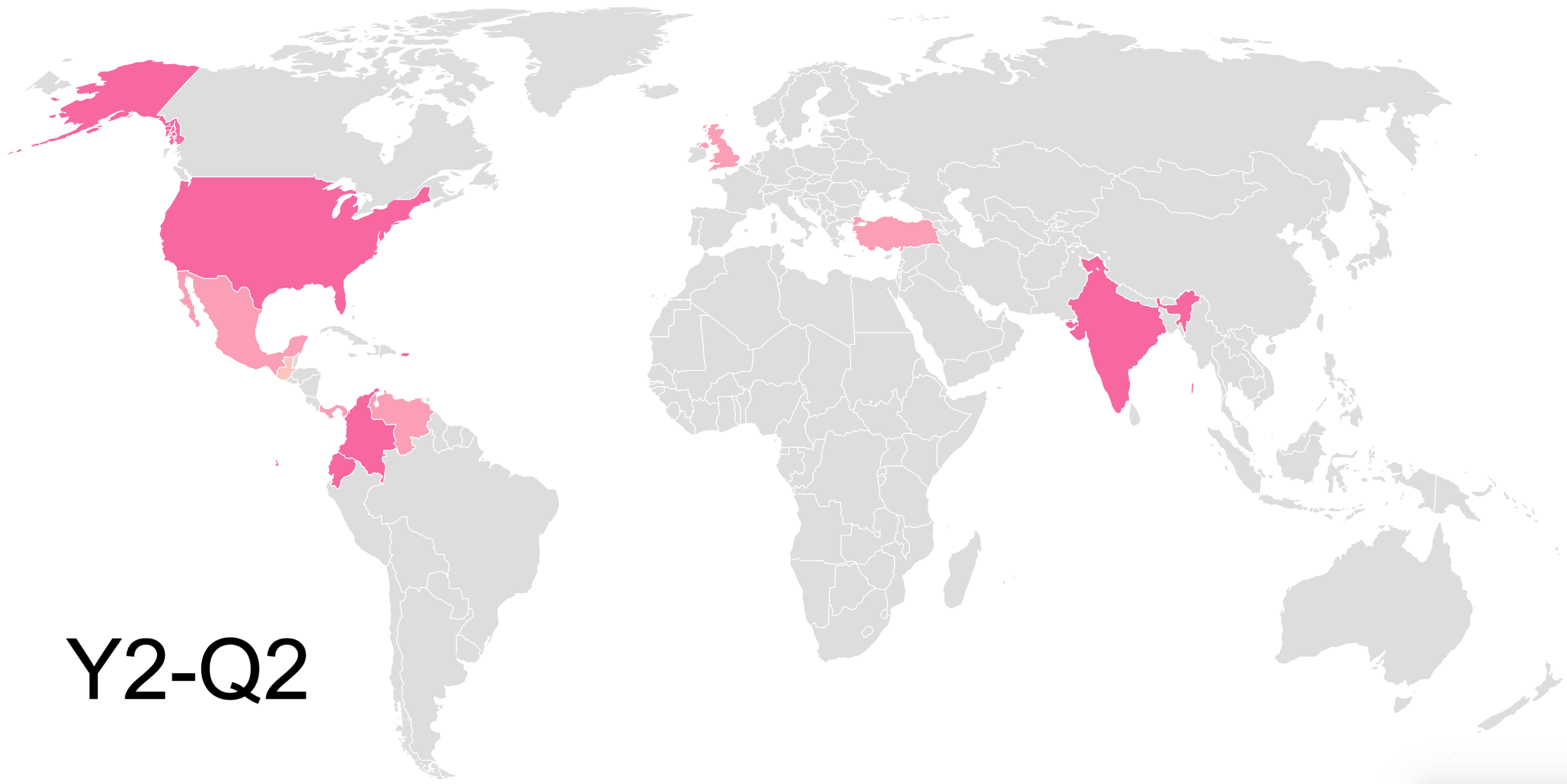}
        \hfill
        \includegraphics[width=0.24\columnwidth]{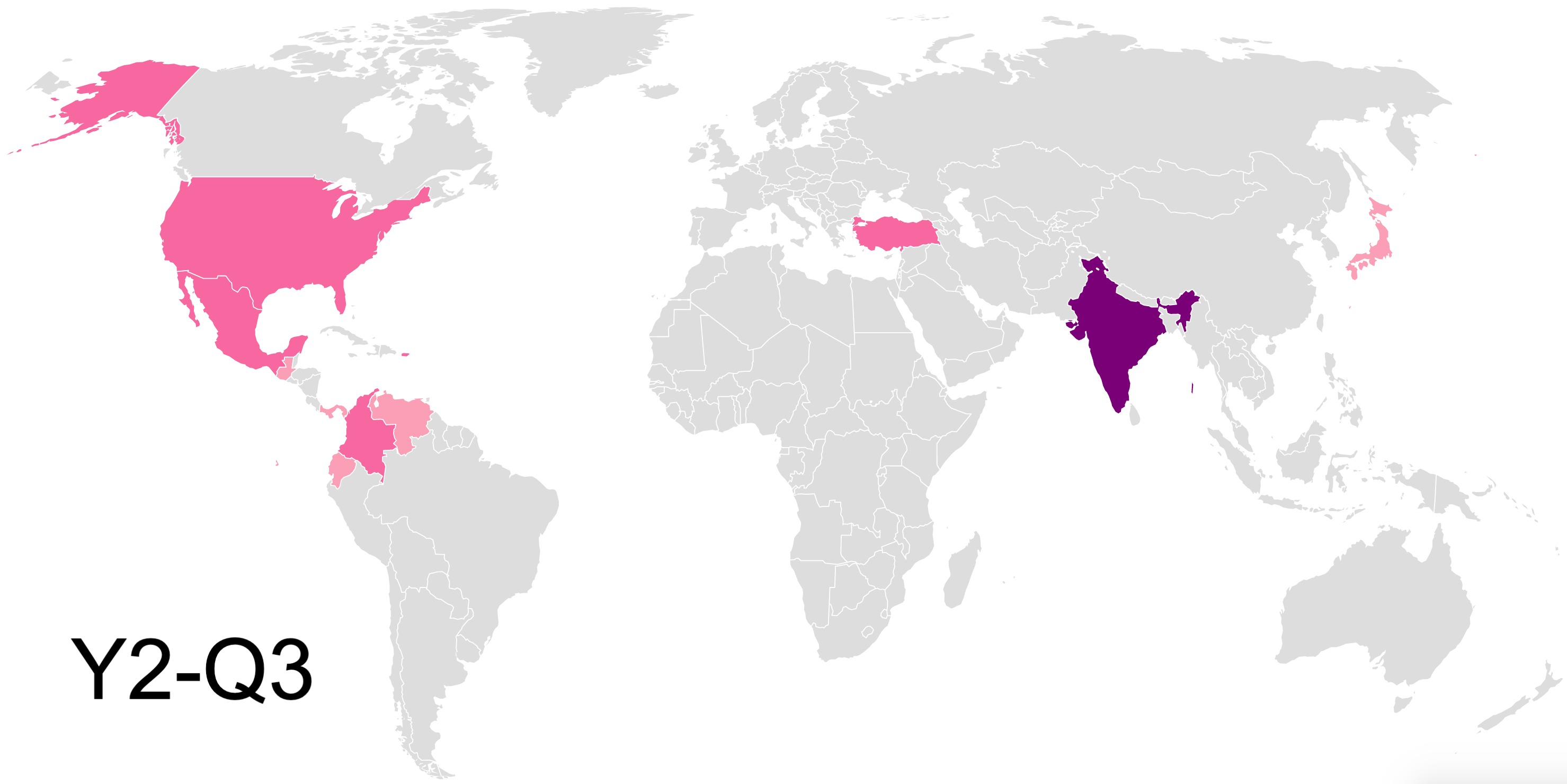}
        \hfill
        \includegraphics[width=0.24\columnwidth]{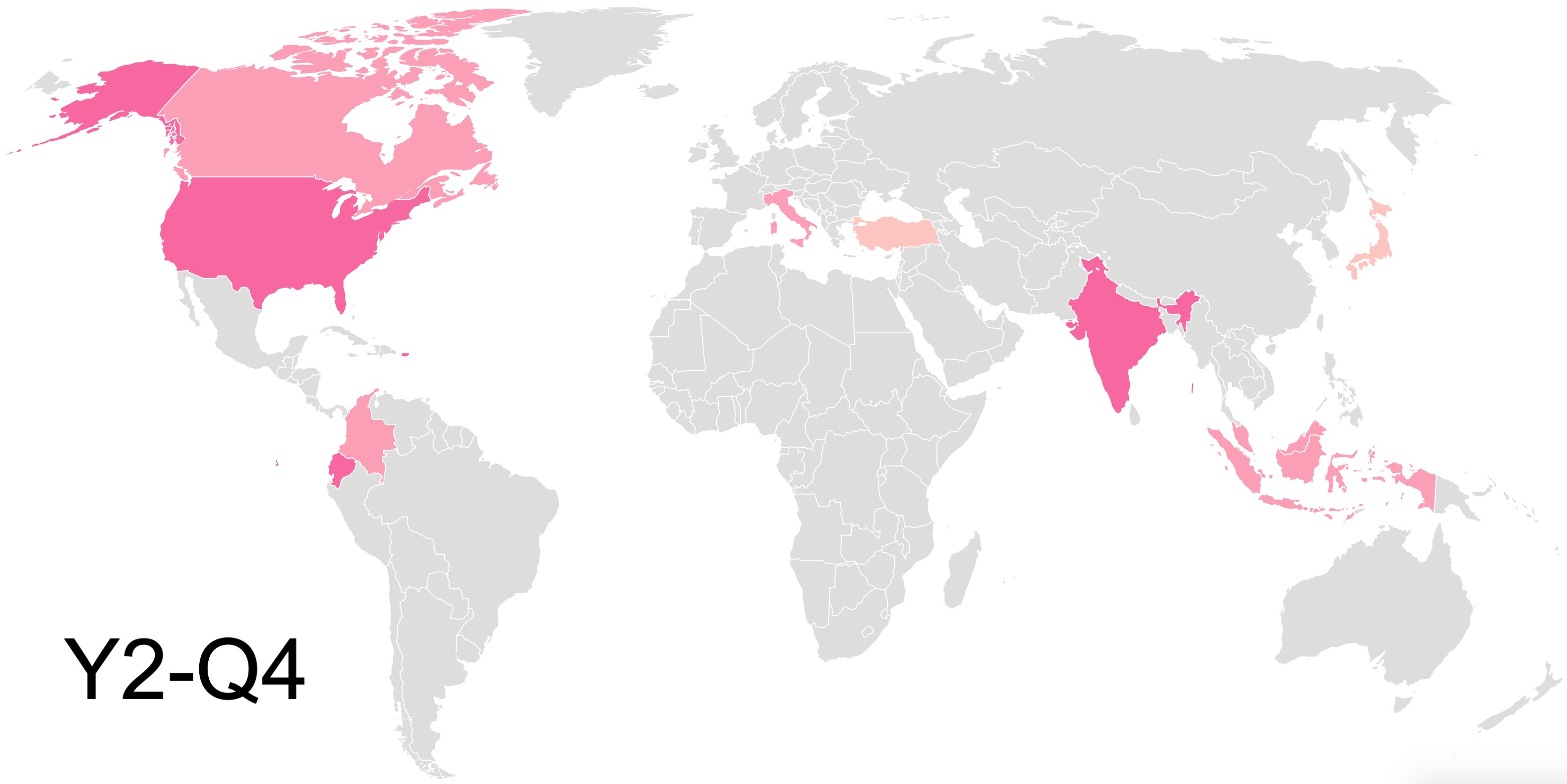}
        \hfill
        \includegraphics[width=0.48\columnwidth]{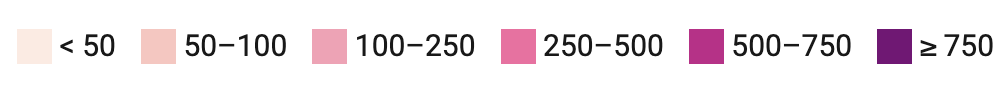}
    \end{minipage}
    \caption[Top-10 countries with the highest landslide reports in each quarter]
    {Top-10 countries with the highest landslide reports in each quarter} 
    \label{fig:quarterly_maps}
    \vspace{-3mm}
\end{figure}

Since its deployment until December 31, 2021, the system has collected more than 54 million tweets and 15 million image URLs, out of which about 2.5 million were deemed unique and downloaded for further analysis. 
Fig.~\ref{fig:weekly_dist_raw_filt} depicts the weekly volume of raw tweets and images collected during this time period as well as the distributions of images filtered by the Junk Filter, Duplicate Filter, and Landslide Detector. The data do not show any gaps, which is an important factor for robust monitoring of real-world events continuously. On average, the Junk Filter eliminates around 76\% of the collected images, the Duplicate Filter further reduces the redundancy by an additional 9\%, and finally, the Landslide Detector classifies only 0.84\% of the remaining 15\% images as landslides. This corresponds to a significant (i.e., more than 99\%) reduction of information overload for our end users. 6,523 of all the detected landslide reports were shared by personal accounts and 4,553 by organizational accounts.
Fig.~\ref{fig:global_map} shows the worldwide distribution of the detected landslide reports while Fig.~\ref{fig:quarterly_maps} highlights the top-10 countries with the highest number of landslide reports in each quarter. We see that US, Ecuador, Colombia, and India experience significant landslide numbers all year round. For India, landslides become even more prevalent in Q3. Likewise, Mexico experiences a significant increase in Q3. In contrast, prominent landslide numbers in Indonesia and Malaysia happen in Q1 and Q4 whereas in the UK they occur more in Q1 and Q2. Turkey experiences most landslides in Q1 through Q3.

\subsection{Validation of the Landslide Model Predictions}
Although the system has collected more than 2.5 million images since its deployment in February 2020, there are only about 17,000 images labeled as landslide (or 38,000 images including near-and-exact duplicates), which corresponds to less than 1\% of the total volume. This highlights the difficulty of the task even though a carefully curated set of landslide-related keywords has been used to collect data from Twitter. 
To validate the performance of the landslide model in the real-world deployment, we sampled N=3,600 tweets with images collected by our system. To avoid overburdening our landslide specialists with noisy data as well as to warrant robust statistics, we sampled only from the subset of tweets with images labeled as non-duplicate and relevant. Our landslide specialists then reviewed these images and annotated them with ground-truth landslide/not-landslide labels. Eventually, we compared the machine-predicted labels with expert annotations to evaluate the performance of the landslide model in a real-world scenario. Table~\ref{tab:real-world_validation} summarizes the number of correct (i.e., True Positive (TP) and True Negative (TN)) and incorrect (i.e., False Positive (FP) and False Negative (FN)) predictions as well as the performance scores such as accuracy, precision, recall, F1, and MCC. Overall, we see that the performance of the model in a real-world scenario is comparable to the experimental results (Section~\ref{ssec:exp_model_landslide}).

\begin{table}[t]
\centering
\caption{Validation of landslide model predictions}
\begin{tabular}{ccccc}
\toprule
\textbf{TP} & \textbf{FP} & \textbf{FN} & \textbf{TN} & \textbf{Total} \\
\cmidrule(lr){1-1}\cmidrule(lr){2-2}\cmidrule(lr){3-3}\cmidrule(lr){4-4}\cmidrule(lr){5-5}
123 & 39 & 43 & 3,395 & 3,600 \\
\midrule
\textbf{Accuracy} & \textbf{Precision} & \textbf{Recall} & \textbf{F1} & \textbf{MCC} \\
\cmidrule(lr){1-1}\cmidrule(lr){2-2}\cmidrule(lr){3-3}\cmidrule(lr){4-4}\cmidrule(lr){5-5}
97.72 & 75.93 & 74.10 & 75.00 & 73.81 \\
\bottomrule
\end{tabular}
\label{tab:real-world_validation}
\vspace{-3mm}
\end{table}

\subsection{Comparison with a Text-based Approach}
Text-based landslide detection is a nascent problem that only a couple of studies have addressed so far~\cite{musaev2015fast,musaev2017rex}.
Since these studies did not share their data sets and models,  we do not have any off-the-shelf text-based landslide classification model to use as a baseline in our study. Therefore, we consider an alternative scenario with a \emph{proxy} text classification model based on lexicon (i.e., keyword) matching, which is already implemented in our system. That is, we assume all the retrieved tweets are already labeled as landslide by a hypothetical model. We then use the previously sampled set of tweets with their expert annotations to compute the precision of a lexicon-based text model. Unsurprisingly, we found that the lexicon-based text model achieved only about 5\% precision (i.e., only about 5\% of the tweets retrieved were indeed related to landslides) while the image classification model achieved 76\% precision as reported before.




\section{Conclusion}
\label{sec:conclusion}

In this paper, we presented a system that was developed through an interdisciplinary collaboration between the computer scientists at the Qatar Computing Research Institute (QCRI) and the earthquake and landslide specialists from the European-Mediterranean Seismological Centre (EMSC) and the British Geological Survey (BGS), respectively. The developed system leverages online social media data in real time to identify landslide-related information automatically using state-of-the-art artificial intelligence techniques. The designed system (i) reduces the information overload by eliminating duplicate and irrelevant content, (ii) identifies landslide images, (iii) infers their geolocation, and (iv) categorizes the user type (organization or person) of the account sharing the information. We presented results of our model development as well as system performance evaluation and benchmarking experiments. We demonstrated the system's success with a real-world deployment. We believe that our system can contribute to harvesting of global landslide data and facilitate further landslide research. Furthermore, it can support global landslide susceptibility maps to provide situational awareness and improve emergency response and decision making.

\bibliographystyle{splncs04}
\bibliography{bibliography}
%




\end{document}